%% file: PRE.tex
\documentclass[%
 amsmath,amssymb,
 aps,longbibliography
]{revtex4-1}

\usepackage{graphicx}% Include figure files
\usepackage{dcolumn}% Align table columns on decimal point
\usepackage{bm}% bold math
\newcommand*{\eea}{\end{array}}

\newcommand*{\bme}{\begin{multiequations}}
\newcommand*{\eme}{\end{multiequations}}

\providecommand\bcdot{\boldsymbol{\cdot}}
\renewcommand*{\Omega}{\varOmega}
\renewcommand*{\Sigma}{\varSigma}

\def\squarebox#1{\hbox to #1{\hfill\vbox to #1{\vfill}}}

\newcommand{\defin}{\stackrel{\scriptscriptstyle\triangle}{=}}

\newcommand{\w}{\boldsymbol{\widetilde u}}
\newcommand{\bu}{\boldsymbol{u}}

\newcommand{\B}{\boldsymbol{B}}
\newcommand{\bsigma}{\boldsymbol{\sigma}}

\newcommand{\xx}{\boldsymbol{x}}

\newcommand{\XX}{\boldsymbol{X}}
\newcommand{\yy}{\boldsymbol{y}}

\newcommand{\nab}{\boldsymbol{\nabla}}

\newcommand{\transp}{^{\scriptscriptstyle T}}

\newcommand{\Exp}{\mathbb{E}}

\newcommand{\dif}{{\mathrm{d}}}
\newcommand{\mbs}[1]{\ensuremath{\boldsymbol{#1}}}
\newcommand{\mbf}[1]{\ensuremath{\boldsymbol{#1}}}

\begin{document}

\preprint{APS/123-QED}

\title{A model under location uncertainty to predict the mean velocity in wall bounded flows}% Force line breaks with \\

\author{Beno\^it Pinier}
\author{Etienne M\'emin}%
\affiliation{%
Inria/ IRMAR/ U. Rennes I, Campus universitaire de Beaulieu, 35042 Rennes Cedex, France
}%


\author{Sylvain Laizet}
\affiliation{
Department of Aeronautics, Imperial College London, South Kensington campus, London SW7 2AZ, UK
}%
\author{Roger Lewandowski}
\affiliation{%
 IRMAR/ U. Rennes I, Campus universitaire de Beaulieu, 35042 Rennes Cedex, France
 }%


\date{\today}% It is always \today, today,

\begin{abstract}
To date no satisfying model exists to explain the mean velocity profile within the whole turbulent layer of  canonical wall bounded flows. 
%The difficulty arises in particular from a change of the dynamical regime between the molecular friction driven dynamical regime in the close-wall region and the log-law layer's dynamics dominated by turbulent dissipation. 
We propose a modification of the velocity profile expression that ensues from a recently proposed stochastic  representation of fluid flows dynamics. This modeling, called {\em modeling under location uncertainty} introduces in a rigorous way a subgrid term generalizing the eddy-viscosity assumption and an eddy-induced advection term resulting from turbulence inhomogeneity. This latter term gives rise to a theoretically well-grounded model for the transitional zone between the viscous sublayer and the turbulent sublayer. An expression of the small-scale velocity component is also provided in the viscous zone. Numerical assessment of the results are provided for turbulent boundary layer flows, for pipe flows and channel flows at various Reynolds numbers.
\end{abstract}

\pacs{Valid PACS appear here}% PACS, the Physics and Astronomy
\maketitle

%\tableofcontents

\section{introduction}

The study of mean velocity profiles in wall-bounded flows is generating an intense research effort. Their knowledge  is as a matter of fact an immense  source of information for many industrial applications and in geophysics to model the interface between atmosphere and ocean. Since the seminal work of \cite{Prandtl25} and  \cite{vonKarman30}, the mean velocity profile is known to be linear in a viscous sublayer near the wall and logarithmic within a turbulent sublayer, located before a region of uniform mean profile.  These models have been derived from different theoretical arguments and confirmed in a wide variety of experiments \citep{jimenez2007we,Marusic10, Klewicki10,  Marusic13}. Between the viscous layer and the logarithmic layer, within an interfacial region, referred to as the buffer zone, no robust model is yet available. 

The lack of a model with clear theoretical foundations  for the buffer zone is essentially due to  a change of the dynamical regime between the viscous and turbulent sublayers. In the former, the molecular friction dominates while in the latter it is a large-scale turbulent mixing  dissipation which governs the flow.  A transitional mechanism is necessarily acting in between these two regions.
In this study we show that by taking properly into account the uncertainties associated to the unresolved (turbulent) components of the wall-bounded flows, it is possible to introduce a theoretically well-defined model for the buffer zone. This model is directly associated to a statistical eddy-induced velocity. Such a drift correction corresponds to the so-called  {\em turbophoresis} phenomenon associated with small-scale inhomogeneity, which drives inertial particles toward the regions of low turbulent diffusivity \citep{Reeks83}. It is also akin to the velocity correction introduced for tracer mean transport in oceanic or atmospheric circulation models \citep{Andrews78}. 

The model used in this study is derived from a large-scale stochastic representation recently proposed by \cite{Memin14} and has been applied with success for various turbulent flows \citep{Chandramouli-CF-18,Kadri-CF-17, Resseguier-JFM-17}. The formulation referred to as modeling under location uncertainty (LU) incorporates a random component as a model of the unresolved small-scale velocities. We briefly describe hereafter its principles.

The proposed stochastic principle relies on a decomposition of the Lagrangian velocity  in terms of a large-scale smooth component, $\w$, and a highly oscillating random component
\begin{equation}
\frac{\dif \XX_t}{\dif t} = \w(\XX_t,t) + \bsigma(\XX_t,t) \dot{\B_t}.
\label{sto-flow}
\end{equation}
The first term represents the large-scale velocity component whereas the second one, written (formally) as the time derivative of a  $d$-dimentional Brownian function, $\dot{\B_t}= \dif \B_t/\dif t$, stands for the fast, unresolved, velocity component. The divergence-free random field involved in this equation  is defined  over the fluid domain, $\Omega$, through the kernel $\breve{\bsigma} (.,.,t) $ of the diffusion operator $\bsigma(.,t)$
 \begin{equation}
\forall \xx \in  \Omega, \;\;\;(\bsigma(\xx,t)\mbs f)^i \defin \sum_j\int_\Omega \breve{\sigma}^{ij}(\xx,\yy,t)  f^j (\yy,t) \dif\yy, \;i,j=1,\ldots, d.
\end{equation}
This operator is assumed to be  of finite norm. The covariance of the random turbulent component is defined as
\[
Q_{ij} (\xx,\xx',t,t') = \Exp\bigl( (\bsigma(\xx,t)\dif \B_t)_i (\bsigma(\xx',t')\dif \B_t)_j\bigr) = c_{ij}(\xx,\xx',t)\delta(t-t')\dif t,
\]
and the diagonal of the covariance tensor,  defined as $a_{ij}(\xx,t) = c_{ij}(\xx,\xx,t)$ is of crucial importance in the following. It has the dimension of a diffusion $(m^2/s)$ and plays the role of a {\em generalized} matrix-valued eddy viscosity. 

The rate of change of a scalar quantity $q$ within a volume transported by the random flow (\ref{sto-flow}) provides us a stochastic representation of the Reynolds transport theorem \cite{Memin14}:
\begin{equation}
\dif\int_{V(t)} q \dif\xx= \int_{V(t)} \bigl(\dif_t q + \nab\bcdot (q\w^*) + \bsigma \dif\B_t \bcdot \nab q -\nab\bcdot (\frac{1}{2} \mbs{a} \nab q\bigr) \dif\xx,
\end{equation}
where the first term represents the increment in time of the random scalar $q$, and the effective advection velocity in the second term is defined as:
\begin{equation}
\w^* = \w - \frac{1}{2} \nab \bcdot \mbs a.
\end{equation}
From this expression the evolution of a conserved scalar (with an extensive property) reads immediately: 
\begin{equation}
\dif_t q + \nab\bcdot (q\w^*) + \bsigma \dif\B_t \bcdot \nab q -\nab\bcdot (\frac{1}{2} \mbs{a} \nab q) =0.
\end{equation}
In this stochastic partial differential equation the forth term is a dissipation term (as the variance tensor, $\mbs a$, is semi-definite positive), the third term represents the advection of the scalar quantity by the random velocity component and the modified advection captures the action of the random field inhomogeneity on the transported scalar. Incompressibility conditions for a fluid with a constant density  are  readily derived as:
\begin{subequations}
\begin{align}
& \nab\bcdot \bsigma\dif \B_t =0,\\
&\nab\bcdot \w^*= \nab\bcdot (\w -\frac{1}{2}\nab\bcdot \mbs a) =0.
\end{align}
\end{subequations}
The first condition is intuitive and enforces a divergence free random component, whereas the second constraint imposes a divergence -free condition on the effective advection. This last relation provides a relation between the smooth resolved velocity component and the variance tensor divergence. For homogeneous random fields (such as an isotropic turbulence) this equation  boils down to a classical divergence-free condition on the resolved velocity component (as the variance tensor is in that case constant).  For isochoric flows with variable density as in geophysical fluid dynamics, interested readers may refer to \cite{Resseguier-GAFD-I-17, Resseguier-GAFD-II-17,Resseguier-GAFD-III-17}.

This random Reynolds transport theorem allows us to derive  from the Newton second principle  the following modified Navier-Stokes system of equations  for an incompressible fluid \cite{Memin14, Resseguier-GAFD-I-17}:
\begin{subequations}
\label{LU}
\begin{align}
&\text{\em Momentum equations} \nonumber\\
&\;\;\;\;\partial_t \w  +\biggl( \bigl(\w - \frac{1}{2} (\nab\bcdot \mbs a)\bigr)\bcdot \nab\biggr)\w   - \frac{1}{2} \nab\bcdot \bigl( (\mbs a  \nab)\w\bigr)= - \frac{1}{\rho}\nab p + \nu\nab^2 \w,\\
&\text{\em Pressure random contribution} \nonumber\\
&\;\;\;\;\nab \dif p_t  =\!
 -   \rho (\bsigma \dif \hat\B_t \bcdot \nab)\w  + \nu\nab^2 \bsigma \dif\B_t,\\
 &\text{\em Mass conservation} \nonumber\\
&\;\;\;\;{\nab} \bcdot(\bsigma \dif \B_t ) =0, \;\;{\nab} \bcdot\w -\frac{1}{2}\nab\bcdot(\nab\bcdot \mbf a) = 0.
\end{align}
\end{subequations}
This system, expressed within  the It\^{o} stochastic integral setting,  corresponds to a large-scale description of the flow in which the effect of  the unresolved random turbulent component is explicitly taken into account. The scale separation operated in this system is obtained under the assumption of a smooth in time large scale component, that is well adapted to the context of this study. A similar stochastic framework arising from an Hamiltonian principle  and a Stratonovich stochastic integral has been proposed in \cite{Holm-15} and analysed in \cite{Crisan-Flandoli-Holm-17} and \cite{Cotter-Gottwald-Holm-17}. This framework leads to enstrophy conservation whereas the LU formulation conserves the kinetic energy of a transported scalar \cite{Resseguier-GAFD-I-17}.

A quite intuitive representation results from this formulation. It  includes in a neat theoretical basis both a generalized eddy-viscosity subgrid model, $(\mbs a \nab) \w$, together with a correction of the advection term associated with turbulence inhomogeneity ($\nab\bcdot\mbs a$). These two terms depend on the variance tensor, $\mbs a(\xx)$. The modified advection describes the statistical effect caused on the large scales by the inhomogeneity of the unresolved velocity component. As we will see, this component plays a fundamental role  in the transitional zone of wall-bounded flow. 
The term $\dif p_t$ corresponds to the pressure associated to the random turbulent component, whereas $p$ is the large scale pressure, $\rho$ is the density and $\nu$ the kinematic viscosity. The last constraint stems from mass conservation and imposes a divergence-free effective advection.

In the next section, we recall the  ideal flow conditions pertaining to the classical derivation of the wall-bounded flow mean velocity profile and develop its expression for the model under location uncertainty. In the following, the $x$ direction is the streamwise direction, $y$ the spanwise direction and $z$ the wall normal direction.

\section{Boundary layer and wall laws}
The derivation of the wall laws relies on the following hypothesis: the large-scale component $\w$ is parallel to the wall plane $\{z=0\}$; the large-scale and small-scale velocity components are stationary and depend only on the distance to the wall, $z$; on the wall the whole flow velocity is zero ($\w=0$ and  $\bsigma =0$); the large-scale pressure $p$ is constant. At fixed depth, the random field is homogeneous with a constant variance tensor. This assumption, which considers no particular dependence on the horizontal plane of the variance tensor seems reasonable. 

  The tangential cumulated  friction exerted by the flow on the wall per time interval, $\Delta t$, is expressed from the shear stress at the wall, which according to our model involves a large-scale component and a small-scale random component
  \begin{equation}
S_t= \rho \nu \int_t^{t+\Delta t} \left.\bigl(\frac{\partial \w}{\partial n} \dif t + \frac{\partial \bsigma}{\partial n}\dif \B_t \bigr)\right |_{z=0},
  \end{equation}
Invoking Ito isometry for the Brownian term, we infer that its mean magnitude reads
\begin{equation}
\Exp \|\mbs S_t\|^2 = (\rho \nu)^2 \Delta t \int |\partial_n \w |^2 \dif t +  (\rho \nu)^2 \int tr \bigl(\partial_n \bsigma(\xx)\; \partial_n \bsigma\transp(\xx) \bigr)\dif t. 
\end{equation}
Assuming that both the normal derivative of the diffusion tensor $\partial_n\bsigma$ and  velocity  $\partial_n \w$ are constant, we get
\begin{equation}
\label{S_t}
(\Exp \|\mbs S_t\|^2)^{\frac{1}{2}}= (\rho \nu) \bigr (|\partial_n \w |^2(\Delta t)^2  + \epsilon^2\Delta t\bigr)^{1/2}. 
\end{equation}
In this expression, $\epsilon^2 = tr \bigl(\partial_n \bsigma(\xx)\; \partial_n \bsigma\transp(\xx)\bigr)$ stands for the variance of the small-scale shear stress (where the dimension of the normal derivative is $[\partial_n \bsigma] =  [T]^{-1/2}$).

The friction velocity ${U_{\tau}}\mbs \delta{\w}$ in the streamwise direction, $\mbs \delta{\w}$,  is now defined from the shear stress as
\begin{equation}
\label{fric-vel}
{ U}_{\tau}= \left(\frac{\Exp (\|S\|^2)^{1/2}}{\rho \Delta t}\right)^{1/2} = \left[ \nu\biggl(|\partial_n \w|^2 + \frac{1}{\Delta t}\epsilon^2 \biggr)^{1/2} \right]^{1/2}.
\end{equation}
 It can be checked this quantity scales as a velocity ($\sim {(L^2/T^2)^{1/2} }$) and for a null uncertainty $\epsilon^2=0$, we obtain the usual definition of the friction velocity: $\widetilde{U}_{\tau} = (\nu |\partial_n \w|)^{1/2})$. For a non null uncertainty, we get a modified expression with a deviation from the standard definition depending on $\Delta t$. It is immediate to observe that when $\Delta t \rightarrow \infty$, ${U}_{\tau}\rightarrow {\widetilde U}_{\tau}$. However, for small time interval and large small-scale velocity stress at the wall, the deviation from the standard definition can be important. The friction velocity ${ U}_{\tau} \approx {\widetilde U}_{\tau}$ is recovered only when the shear stress variance is much smaller than the time interval:  $\epsilon^2\ll \Delta t$.
%We introduce now the spatial and temporal dimensional scaling: 
%\begin{equation}
%\lambda_{bl}= \frac{\nu}{U_{\tau}} \text{ and } \tau = \frac{\nu}{U^2_{\tau}},
%\end{equation}
%and consider the scaled velocity profile along the vertical direction:
%\begin{align}
%\mbs L(z') &=  \frac{\w(\lambda_{bl} z')}{U_{\tau}}  + \frac{1}{U_{\tau}} \tau^{-1/2}\bsigma (\lambda_{bl}z')\frac{\dif \B_{ t'}}{\dif t'}.
%\end{align}
%Assuming the profile is universal, we get
%\begin{equation}
%\label{prof}
%\forall z\in [0, z_{\tau}] \;\;\bu(z')  = U_{\tau}\; \mbs L(z').
%\end{equation}
\subsection{Boundary layer structure}
As classically admitted, the boundary layer is formed of two main sublayers: the viscous sublayer and the turbulent layer. The former corresponds to a region of contact between the wall and the fluid, where the flow is  driven mainly by the molecular shear stress. In the latter,  the flow is dominated by the large-scale shear stress associated to the unresolved fluctuations (here the small-scale random field). For the LU flow dynamics  and the TBL ideal configuration described above, the stationary equations for the  mean velocity component in these two sub-layers are described below.
\subsubsection{Viscous sublayer}
In the viscous sublayer, extending from the wall ($z=0$) to a distance ($z=z_0$),  molecular viscosity dominates at all scales.
From system (\ref{LU}), we get that the large-scale drift component exhibits a constant variation depth, while  the small-scale component is necessarily spatially very smooth (harmonic); and the random turbulent pressure diffusion term is consequently constant:
\begin{subequations}
\label{sto-inc-NS}
\begin{align}
&\text{\em Large-scale component}\nonumber\\
&\;\;\;\;\nu \nab^2 \w =0 \Rightarrow \partial_z \w = C_1 \label{u-tilde-Cte},\\
&\text{\em Small scale component}\nonumber\\
&\;\;\;\;\nu \nab^2 \bsigma \dif \B_t \!=\!0 \Rightarrow\nab^2 \!\bsigma^{ij} \!=\!0\label{harmo-cond},\\
&\text{\em Turbulent pressure }\nonumber\\
&\;\;\;\;\dif p_t = C _2\label{TP},\\
&\text{\em Incompressibility }\nonumber\\
&\;\;\;\;{\nab} \bcdot(\bsigma \dif \B_t ) =0. \label{Hincomp}
\end{align}
\end{subequations}
\subsubsection{Turbulent sublayer}
In the turbulent sublayer, delimited between the end of the viscous layer $z=z_0$ and an upper limit $z=z_1$, the dynamics is driven by the combination of the large-scale diffusion and the molecular friction. From the ideal TBL assumptions, for this sublayer, system (\ref{LU}) reads 
\begin{subequations}
\begin{align}
&\text{\em Large-scale component }\nonumber\\ 
&-\partial_z a_{zz} \partial_z \w - \partial_z \bigl((a_{zz} + 2 \nu)\partial_z \w \bigr) =0 \label{eq-LS},\\
&\text{\em Turbulent pressure horizontal gradients }\nonumber\\
&\nab_H \dif p_t \!= \!\partial_z \w (\bsigma\dif\B_t)_z \!+\! \nu \nab^2 (\bsigma \dif \B_t)_H =0,\\
&\text{\em Turbulent pressure vertical gradient }\nonumber\\
&\partial_z \dif p_t = \nu\nab^2(\bsigma \dif \B_t)_z \label{dzp},\\
&\text{\em Incompressibility } \nonumber\\
&{\nab} \bcdot(\bsigma \dif \B_t ) =0,  \;\; \nab\bcdot(\nab\bcdot \mbf a) = 0. \label{incompressibility}
\end{align}
\end{subequations}
We observe that the modified advection term  caused by an eventual inhomogeneity of the turbulence is also involved in equation (\ref{eq-LS}). The second term of this equation represents the diffusion due to the molecular friction and the small-scale mixing activity. 

 From these two systems,  the expressions of the mean velocity profile can be inferred for both regions.   We will start first by the viscous sublayer.
\subsection{Velocity expression in the viscous sublayer}
At the interface between the viscous and turbulent sublayers $(z=z_0)$, the large-scale normal derivative of the velocity being constant, a null advection of $\w$ by the random field ($\partial_z \w (\bsigma \dif \B_t)_z =0$) in the turbulent pressure equation \eqref{TP}  implies that $ \left. (\bsigma \dif \B_t)_z\right |_{z=z_0} =0$. The null boundary condition  of the random field at the wall and the harmonic condition (\ref{harmo-cond}) together with  the strong maximum principle indicates that the  turbulent component is necessarily  a 2D $(i.e. (\bsigma \dif \B_t)_z =0)$ incompressible \eqref{Hincomp} flow everywhere in the viscous layer. Note this 2D flow is not constant on the viscous sublayer volume as its horizontal components depends on depth.
Integrating over the viscous layer depth the harmonic condition (\ref{harmo-cond}) of this horizontal random field $(\nab^2_H (\bsigma \dif \B_t)_H = - \partial^2_{zz} (\bsigma \dif \B_t)_H)$ -- with the subscript $H$ denoting the horizontal components -- we get
  \[
\nab^2_H \int_0^{z_0} (\bsigma \dif \B_t)_H = - \left.\partial_{z} (\bsigma \dif \B_t)_H \right|_{z=z_0} + (\partial_n\bsigma\B_t)_H.
 \]
 The left-hand side term corresponds to an empirical mean along the vertical direction. Since it is a zero-mean random variable, it tends to zero (discretizing the interval with enough points) and thus
 \[
 \left.\partial_{z} (\bsigma \dif \B_t)_H \right|_{z=z_0} = (\partial_n\bsigma\B_t)_H.
 \]
 The right-hand term is an homogeneous random field with variance $\epsilon^2 \dif t$. The left-hand side random field has the same characteristics. Again due to the harmonic constraint and the strong maximum principle, its variance increases linearly with $z$. Therefore, the whole random field  can be defined from a  unitary 2D  divergence-free  Gaussian random field, $\mbf \eta_t$, on the whole viscous layer  as
 \begin{equation}
 \label{2D-viscous-MRF}
 (\bsigma (\xx)\dif\B_t)  =  \epsilon z  \sqrt{\dif t} \;\mbs \eta (\xx), \;\forall z \in [0,z_0].
 \end{equation}
This allows us to state: 
 {\em in the viscous sublayer the small-scale  component is a 2D  divergence free random field characterized by a variance, which depends on the wall shear stress variance with a linear increase in time and a growth in the viscous layer of the square of the depth. Its vorticity is slanted, $\nab\times\bsigma (\xx)\dif\B_t = \epsilon\sqrt {\dif t}(-\eta_y, \eta_x, z (\partial_x \eta_y -\partial_y \eta_x))\transp$ and its mean magnitude intensifies linearly with the distance to the wall. As a result, it forms  curved cones of vorticity}. 
 
Besides, from the friction velocity definition (\ref{fric-vel}), (\ref{u-tilde-Cte}) and because $\partial_z \w = \partial_n \w = C_1$  we have (with $\partial_z \w > 0 $) 
\begin{equation}
\partial_z \w =  \bigl (  \frac{1}{\nu^2} U^4_{\tau} -\frac{1}{\Delta t}\epsilon^2 \bigr)^{1/2} \mbs\delta\w = \frac{1}{\nu}\widetilde{U}^2_{\tau} \mbs\delta\w, 
\end{equation}
where, $\widetilde{U}_{\tau} = (\nu |\partial_n \w |)^{1/2}$, stands for the friction velocity associated to the large-scale component. 
Integrating along $z$ and since $\w(0)=0$, we therefore get
\begin{align}
\w(z) &= \frac{1}{\nu}\widetilde{U}^2_{\tau} z \;\mbs\delta\w.\label{viscous-drift}
\end{align}
Gathering the large-scale and small-scale components, the whole infinitesimal displacement field over time interval, $\Delta t$, in the viscous layer finally reads
\begin{align}
\forall z\in [0, z_0] \;\;\bu(z)\Delta t  &=  \frac{1}{\nu}\widetilde{U}^2_{\tau}  z\; \mbs\delta\w \Delta t+   \epsilon z (\Delta t)^{1/2} \mbs \eta .
\end{align}
The small-scale zero-mean random component has a variance  $V=  \epsilon^2  z^2 \Delta t$. The mean velocity profile is given by \eqref{viscous-drift} and the usual linear expression is retrieved. It is interesting to note that this profile can be specified from the friction velocity associated to the long-time average velocity field (if it can be computed), or from \eqref{fric-vel} if only smooth velocity snapshots on a given period of time together with an estimation of the small-scale shear stress variance are available. This latter case corresponds to the situation often encountered for the study of oceanic or atmospheric flows.

\subsection{Velocity expression in the buffer and turbulent sublayer}
   
In the vicinity of the viscous sublayer the molecular friction still dominates whereas at the end of the turbulent sublayer the large-scale shear stress is predominant. This stress depends directly on the small-scale variance. We assume that at the end of the turbulent sublayer the wall has no influence on  the turbulence. According to this hypothesis, the variance tensor tends toward an expression that does not depend on $z$ anymore. As a consequence, at the end of the turbulent layer 
 the dynamics  of the large-scale component (\ref{eq-LS}) corresponds to  an eddy-viscosity formulation
 \begin{equation}
(a_{zz}+ 2\nu)\partial^2_{zz} \w =0,
\end{equation}
which leads  to the logarithmic profile. The log law is however known to poorly fits the transitional buffer region coming into play just after the viscous layer. For that reason, we choose to separate the turbulent layer into a logarithmic region and a transitional buffer zone.
\subsubsection{Buffer zone}
In the buffer zone, delimited by the end of the viscous layer, $z_0$, and the beginning of the logarithmic region, $z_L$, we assume a strict independence of the variance tensor with respect to the horizontal directions. With this assumption, the small-scale component is a 3D homogeneous random field at a fixed given depth. In other words, the  variance tensor, $\mbs a$, depends only on depth. From the incompressibility  constraint (\ref{incompressibility}) we get
\begin{equation}
\nab\bcdot \nab \bcdot  \mbs a =0 \implies  \partial^2_{zz} a =0 \implies \partial_z a_{zz} =C'.
\end{equation}
At the interface  $a_{zz}(z_0)=0$, which yields $a_{zz}(z) = C'(z-z_0)$ where $ C'$ scales as a velocity. From similarity principles, it is natural to define
\begin{equation}
a_{zz}(z) = \widetilde{\kappa} \widetilde{U}_{\tau}(z-z_0),
\label{sim}
\end{equation}
where $\widetilde{\kappa}$ denotes a constant. This constant is related to the slope of the variance tensor coefficient along $z$. This constant is completely different from the von Karman constant, attached to the logarithmic region.  We nevertheless designate it with the same letter to refer to the parameter of the classical wall law models. 

At the interface $z=z_0$, let us recall that we have from (\ref{viscous-drift})
\begin{align}
 \partial_z \w(z_0) =\frac{1}{\nu}\widetilde{U}^2_{\tau}  \mbs\delta\w,
\label{eq:derivUz0}
\end{align}
and a null value for the vertical variance tensor $(a_{zz} =0)$. Integrating (\ref{eq-LS}) with the above boundary condition (\ref{eq:derivUz0}) gives an expression for $\partial_z\w$. A second integration of the same equation gives  the following  velocity profile within the buffer zone
\begin{equation}
\forall z\in [z_0,z_L]\;\;\w(z)  =  \w(z_0)  -   \widetilde{U}_{\tau} \frac{4\nu}{\widetilde{\kappa}}\left( \frac{1}{\widetilde{\kappa} \widetilde{U}_{\tau} (z-z_0)+ 2 \nu } - \frac{1}{2 \nu}\right) \delta\w.
\label{eq:deriv3}
\end{equation}
It can be checked that   $\w(z)$ and $\partial_z \w(z)$ are indeed positive and therefore verify the fundamental properties of the large-scale velocity  in the TBL. The buffer zone is restricted to an area located between the end of the viscous zone (at $z=z_0$) and the beginning of the logarithmic region (at $z=z_L$).
\subsubsection{Logarithmic region} To reach a logarithmic profile from two successive integrations of (\ref{eq-LS}) the variance tensor cannot be linear anymore. It necessarily scales as the square-root of the wall distance ($a_{zz}\sim\sqrt z$).  Because the flow is continuous in the whole turbulent boundary layer, we get the following expression
\begin{equation}
a_{zz}(z) = \widetilde{\kappa} \widetilde{U}_{\tau} (z_L-z_0)\sqrt{ \frac{z}{z_L}}, \;\;\forall z\in [z_L, z_1],
\label{eq:Azz_sqrt}
\end{equation}
for the wall-normal variance tensor value. To satisfy the incompressibility condition (\ref{incompressibility}), such an expression comes to relax the strict independence on $x$ and $y$ of the variance tensor (i.e. $\partial^2_{z^2} a_{zz}\neq 0$). This is coherent with the apparition in the flow of elongated structures such as streaks (see \cite{Jimenez-JFM-18} for a recent and complete review on the subject).
 For the region located between the buffer zone limit ($z_L$) and the end of the turbulent sublayer ($z_1$), we have
\begin{equation}
\w(z) = \w(z_L) + \frac{\partial_z \w(z_L)  z_L }{\widetilde{\kappa} \widetilde{U}_{\tau}(z_L-z_0)} \ln \left(\frac{z}{z_L}\right)\delta\w.
\label{eq:drift_azz_sqrt}
\end{equation}
This profile differs slightly from the usual logarithmic law. In particular, we notice that the von Karman constant, which weights the usual log-law, has here a more complex expression that depends among other things on the separation limit between the buffer zone and the logarithmic region. 
 
 \noindent{\bf Summary} The location uncertainty principle allows us to formalize a continuous model for the mean velocity profile within the whole turbulent layer of an ideal wall bounded flow. Due to a modification of the advection velocity induced by the turbulence inhomogeneity, the new model enables to connect on a firmed basis the two classical velocity profiles in the viscous and logarithmic sublayers.  The velocity profile in this transitional so-called buffer zone scales as $1/z$. This mean velocity profile depends on a matrix valued eddy viscosity function, called the variance tensor, and is related to the variance of the (random) velocity fluctuations. The value of this tensor in the wall  normal direction, $a_{zz}$, is the main driver of this model. It scales as $a_{zz} \propto 0$; $a_{zz}\propto z$; and $a_{zz} \propto \sqrt{z}$ in the viscous, buffer and logarithmic zones, respectively.The whole profile depends on the usual friction velocity defined either from  the average velocity field, or from shorter-time averages and an estimation of the small-scale shear stress variance. It depends also on three parameters which are the depth of the viscous layer, the depth of  the logarithmic zone start (or the end of the buffer sublayer) and a constant related to the slope of the wall-normal eddy viscosity $a_{zz}$ in the buffer zone.

\section{Numerical validation}

This section is devoted to the numerical assessments  of the theoretical expressions derived in the previous section. We will in particular assess the mean velocity profile of several wall bounded flows going from moderate Reynolds numbers to high Reynolds numbers. The flows considered will be either turbulent boundary layer flows, pipe flows or channel flows whose results have been well documented in the literature and for which data are available.

The data come from three different databases. The turbulent boundary layer simulations are provided by the Universidad Politecnica de Madrid and are described in  \cite{sillero2013one,Simens2009,Borrell2013}. These simulations lies in a range $Re_\theta \in [2780-6650]$ and an equivalent Reynolds friction number $Re_\tau \in [1000,2000]$. The adimensional Reynold number $Re_\theta = U_\infty \theta/\nu$ is based on the free-stream velocity $U_\infty$ and the momentum thickness $\theta=\int_0^\infty U/U_\infty(1- U/U_\infty )\dif y$. The Reynolds friction number  $Re_\tau =U_{\tau} \delta/\nu$ , is based on the kinematic viscosity, the friction velocity, and the flow thickness $\delta$, which is taken to be the half-width in channels, the 99\% thickness in boundary layers, and the radius in pipes. 

The velocity profiles of the pipe flow simulations are provided by the Royal Institute of Technology of Stockolm (KTH). We have used data at $Re_\tau = 180$, 360, 550, and 1000. These  simulations are  described and analyzed  in \cite{elKhoury2013}. As a last example, we also considered a channel flow for a very high Reynolds number case ($Re_\tau \approx 5200$) provided by  \cite{lee2015direct}.

For all these flows, the limit of the viscous zone ($z_0$), the constant $\widetilde{\kappa}$ and the limit of the buffer zone are estimated through a least-squares procedure and a gradient descent optimization. The least squares cost function between the data and the model is expressed on a section going from the wall to the end of the logarithmic section. The initial value of the parameter are manually fixed from a first rough profile. 

We show in the following, the results obtained for the three types of flows mentioned previously. 
\subsection{Turbulent boundary layer}
We first gather in table \ref{tab:TBL}  the optimal triplet of parameters associated to the reconstructed mean velocity profiles. As it can be observed the values of these parameters vary differently. As classically observed, the limit of the viscous zone lies in a tight range around  $z^+_0 = 5$. The end of the buffer zone, in the other hand, varies more significantly when the Reynolds number is increased. The constant, $ \widetilde \kappa$, -- which corresponds to the slope of the variance tensor coefficient (along $z$)  in the buffer zone -- , is relatively constant and lies within a range between $0.157$ and $0.161$.   
%\begin{table}
%\begin{equation }
%\begin{array}{|c |c| c| c|}
%\hline
%Re_* & z_0^+ & z_l^+ &  \kappa \\
%\hline
%1306 & 4.94& 48.22 & 0.158  \\
%\hline
%1437 & 4.97 & 48.29 & 0.157 \\
%\hline
%1709 &  5.08 & 49.01  & 0.161\\
%\hline
%1989 &  4.8976 & 50.38  & 0.15776 \\
%\hline
%\end{array}
%\label{tab:Variable_TBL}
%\end{equation}
%\caption{Parameters value of the mean velocity profile of turbulent boundary layer flows at }
%\end{table}

\begin{table}
\caption{Parameters value of the mean velocity profile for turbulent boundary layer flows.}
\begin{ruledtabular}
\begin{tabular}{lccc}
$Re_\tau$ &$ z_0^+$ & $z_L^+ $& $ \widetilde\kappa $\\
\hline
1306 & 4.94& 48.22 & 0.158  \\
%\hline
1437 & 4.97 & 48.29 & 0.157 \\
%\hline
1709 &  5.08 & 49.01  & 0.161\\
%\hline
1989 &  4.90 & 50.38  & 0.158\\
%\colrule
\end{tabular}
\end{ruledtabular}
\label{tab:TBL}
\end{table}

\begin{figure}
\input{UMean_TBL_1300.tex}
\caption{Velocity profiles for the turbulent boundary layer at $Re_\tau = 1306$. The green curve is the DNS reference velocity profile; the blue dot lines show the classical laws (linear then logarithmic) and the red dots corresponds to the profile of the model under location uncertainty. }
\label{TBL-1306}
\input{UMean_TBL_1437.tex}
\caption{Velocity profiles for the turbulent boundary layer at $Re_\tau = 1437$. The green curve is the DNS reference velocity profile; the blue dot lines show the classical laws (linear then logarithmic) and the red dots corresponds to the profile of the model under location uncertainty. }
\label{TBL-1437}
\end{figure}

\begin{figure}
\input{UMean_TBL_1709.tex}
\caption{Velocity profiles for the turbulent boundary layer at $Re_\tau = 1709$. The green curve is the DNS reference velocity profile; the blue dot lines show the classical laws (linear then logarithmic) and the red dots corresponds to the profile of the model under location uncertainty. }
\label{TBL-1709}
\input{UMean_TBL_1989.tex}
\caption{Velocity profiles for the turbulent boundary layer at $Re_\tau = 1989$. The green curve is the DNS reference velocity profile; the blue dot lines show the classical laws (linear then logarithmic) and the red dots corresponds to the profile of the model under location uncertainty. }
\label{TBL-1989}
\end{figure}

The profiles of the new model are compared to the classical wall laws: $\w = z\widetilde{U}_{\tau}^2/ \nu$ in the viscous sublayer, and $\w = u_{\tau} \left( \frac{1}{\kappa} \ln (z \frac{\widetilde{U}_{\tau}}{ \nu}) + B \right)$ above the viscous sublayer, where the constants $\kappa$ and $B$ are optimally set from the data. These results are shown in figures \ref {TBL-1306}--\ref{TBL-1989} for $Re_\tau=1306$, $1437$, $1709$, and $1989$ respectively. 

In the viscous sublayer, there is no difference between the classical models and the new one \eqref{viscous-drift}. Both of them perfectly superimpose with the reference data. In that region, the variance tensor along the wall-normal direction, $a_{zz}$, is zero. 

In the buffer region,  where $a_{zz}(z)$ is linear, there is no theoretically well grounded model available for the velocity profile. In this transition area both the classical linear profile or the logarithmic profile deviate significantly from the reference profiles. At the opposite,  the new model fits well the data t all the Reynolds number investigated. The LU model, unlike adhoc formulation \citep{Spalding1961}, enables us to devise a physically coherent model, in which a modified advection ensuing from the unresolved velocity inhomogeneity plays a fondamental role. 

In the logarithmic region, both the classical log-law model and the LU model perform similarly and provide very good results for $Re_\tau=1306$ and $1437$. For higher Reynolds number ($Re_\tau=1709$, and $1989$) they both still match very the data; however the LU log profile approaches even better the data. The good performance of the LU mean velocity profile support  the validity of the wall-normal variance tensor profile in the turbulent layer (e.g. linear and square-root profiles of $a_{zz}(z)$ in the buffer zone and in logarithmic layer respectively) as the velocity profile highly depends on $\partial_z a_{zz}\partial_z \w(z_l)$ \eqref{eq-LS}. 

\subsection{Pipe Flow}

Let us now examine the results obtained for the pipe flows data. The optimal parameters are shown in table \ref{tab:PIPE}. We observe that the value of  the viscous layer thickness ($z_0^+$) decreases with the increase of the friction Reynolds number, while the limit of the buffer zone ($z_L^+$) grows; the constant $\widetilde\kappa$ remains almost constant except for the first simulation associated to the lowest friction Reynolds number.

The mean velocity profiles are plotted in figures \ref{fig:PipeFlow_180},\ref{fig:PipeFlow_360}, \ref{fig:PipeFlow_590},  and \ref{fig:PipeFlow_1000}. As previously, the modified LU model is shown in red, and is compared to the DNS data and the classical logarithmic and viscous velocity profiles. Let us note that compared to turbulent boundary layer flows, pipe flows exhibits a much shorter layer with a logarithmic profile. In the same way as in the turbulent boundary layer case, we clearly see that the model proposed is almost perfectly in agreement with the data up to the end of the logarithmic layer. This model is again particularly relevant for the buffer zone, where no physical model has been yet derived in the literature from a classical eddy-viscosity concept.    

%\begin{equation}
%\begin{array}{|c |c| c| c|}
%\hline
%Re_* & z_0^+ & z_l^+ &  \kappa \\
%\hline
%180 & 5.61 & 43.754 & 0.15  \\
%\hline
%360 & 5.27 & 43.85 & 0.158\\
%\hline
%590 &  5.11 & 46.31  & 0.158\\
%\hline
%1000 &  5.05 & 49  & 0.158 \\
%\hline
%\end{array}
%\label{tab:Variable_PipeFlow}
%\end{equation}
\begin{table}
\caption{Parameters value of the mean velocity profile for pipe flows}
\begin{ruledtabular}
\begin{tabular}{lccc}
$Re_\tau$ &$ z_0^+$ & $z_L^+ $& $  \widetilde\kappa $\\
\hline
180 & 5.61 & 43.75 & 0.150  \\
360 & 5.27 & 43.85 & 0.158\\
550 &  5.11 & 46.31  & 0.158\\
1000 &  5.05 & 49.00  & 0.158 \\
%\colrule
\end{tabular}
\end{ruledtabular}
\label{tab:PIPE}
\end{table}

\begin{figure}
\input{UMean_Pipe_180.tex}
\caption{Velocity profiles for the pipe flow at $Re_\tau = 180$. The green curve is the DNS reference velocity profile; the blue dot lines show the classical laws (linear then logarithmic) and the red dots corresponds to the profile of the model under location uncertainty. }
\label{fig:PipeFlow_180}
\input{UMean_Pipe_360.tex}
\caption{Velocity profiles for the pipe flow at $Re_\tau = 360$. The green curves is the DNS reference velocity profile; the blue dot lines show the classical laws (linear then logarithmic) and the red dots corresponds to the profile of the model under location uncertainty. }
\label{fig:PipeFlow_360}
\end{figure}

\begin{figure}
\input{UMean_Pipe_550.tex}
\caption{Velocity profiles for the pipe flow at $Re_\tau = 550$. The green curve is the DNS reference velocity profile; the blue dot lines show the classical laws (linear then logarithmic) and the red dots corresponds to the profile of the model under location uncertainty. }
\label{fig:PipeFlow_590}
\input{UMean_Pipe_1000.tex}
\caption{Velocity profiles for the pipe flow at $Re_\tau = 1000$. The green curve is the DNS reference velocity profile; the blue dot lines show the classical laws (linear then logarithmic) and the red dots corresponds to the profile of the model under location uncertainty. }
\label{fig:PipeFlow_1000}
\end{figure}

\subsection{Channel flow}
The last example concerns the channel flow at $Re_\tau= 5200$ of \cite{lee2015direct}. For this flow the parameters estimated from the data are gathered in table \ref{tab:CHANNEL}.
\begin{table}
\caption{Parameters value of the mean velocity profile for a channel  flow}
\begin{ruledtabular}
\begin{tabular}{lccc}
$Re_\tau$ &$ z_0^+$ & $z_L^+ $& $  \widetilde\kappa $\\
\hline
5200 & 5.0 & 45.0 & 0.16  \\
%\colrule
\end{tabular}
\end{ruledtabular}
\label{tab:CHANNEL}
\end{table}
As it can be observed they are sensibly the same with a shorter buffer region than in the previous examples.
\begin{figure}
\input{UMean_5200.tex}
\caption{Mean velocity profiles for the channels flow at $Re_{\tau} \approx 5200$ \cite{lee2015direct}. The green curve is the DNS reference velocity profile; the blue dot lines show the classical laws (linear then logarithmic) and the red dots corresponds to the profile of the model under location uncertainty. }
\label{fig:Wall_Law_uncertainty_Drift}
\end{figure}
The plot of the mean velocity profiles is shown figure \ref{fig:Wall_Law_uncertainty_Drift}. As previously, the LU model explains very well the data. In the buffer zone the LU model almost perfectly match the data. In the the logarithmic region the LU  logarithmic law and the classical one perform identically.

\section{Conclusion}
The modeling under location uncertainty provides a novel  expression of the wall-law velocity profiles. For three different wall-bounded flow configurations, going from moderate to high Reynolds numbers, it is shown that the LU model allows us to devise a physical model for the mean velocity profile that matches very well the data in the whole turbulent boundary layer while remaining  continuous and differentiable. The derivation unveils in particular the role played by the advection correction due to turbulence  inhomogeneity in the transitional zone between the viscous and the logarithmic sublayers. This model is very promising as it provides -- as far as we know -- for the first time a physically relevant model for the buffer region. The LU derivation relies on a random modeling of the fluctuation velocity fields and exhibits a tensor playing the role of a matrix valued eddy-viscosity term generalizing de facto the Boussinesq assumption and the Prandlt mixing length. This tensor corresponds to the fluctuation variance times a decorrelation time; its wall-normal component is null in the viscous zone, linear in the buffer zone and scales as $\sqrt{z^+}$ in  the logarithmic layer. The inhomogeneity of this tensor enforces a modified advection term which can be interpreted as the statistical influence of the fluctuation inhomogeneity on the large-scale advection component. This modification has a major contribution in the transitional buffer zone. The new velocity profile associated to the LU model relies principally on a new parameter related to the slope of the variance tensor wall-normal  component in the buffer zone. 

This new model opens very exciting perspectives for the set up of models  ``\`a la'' Monin-Obukhov for fluid flows with thermal stratification as well as the establishment of more accurate wall functions for Large Eddy Simulation. It can also be promising for industrial problems involving pipe flows. 

%\bibliographystyle{}
 % \begin{acknowledgments}
  %The author wish to acknowledge \end{acknowledgments}

\bibliography{glob.bib}% Produces the bibliography via BibTeX.

\end{document}

%% file: UMean_TBL_1300.tex
% GNUPLOT: LaTeX picture with Postscript
\begingroup
  \makeatletter
  \providecommand\color[2][]{%
    \GenericError{(gnuplot) \space\space\space\@spaces}{%
      Package color not loaded in conjunction with
      terminal option `colourtext'%
    }{See the gnuplot documentation for explanation.%
    }{Either use 'blacktext' in gnuplot or load the package
      color.sty in LaTeX.}%
    \renewcommand\color[2][]{}%
  }%
  \providecommand\includegraphics[2][]{%
    \GenericError{(gnuplot) \space\space\space\@spaces}{%
      Package graphicx or graphics not loaded%
    }{See the gnuplot documentation for explanation.%
    }{The gnuplot epslatex terminal needs graphicx.sty or graphics.sty.}%
    \renewcommand\includegraphics[2][]{}%
  }%
  \providecommand\rotatebox[2]{#2}%
  \@ifundefined{ifGPcolor}{%
    \newif\ifGPcolor
    \GPcolorfalse
  }{}%
  \@ifundefined{ifGPblacktext}{%
    \newif\ifGPblacktext
    \GPblacktexttrue
  }{}%
  % define a \g@addto@macro without @ in the name:
  \let\gplgaddtomacro\g@addto@macro
  % define empty templates for all commands taking text:
  \gdef\gplbacktext{}%
  \gdef\gplfronttext{}%
  \makeatother
  \ifGPblacktext
    % no textcolor at all
    \def\colorrgb#1{}%
    \def\colorgray#1{}%
  \else
    % gray or color?
    \ifGPcolor
      \def\colorrgb#1{\color[rgb]{#1}}%
      \def\colorgray#1{\color[gray]{#1}}%
      \expandafter\def\csname LTw\endcsname{\color{white}}%
      \expandafter\def\csname LTb\endcsname{\color{black}}%
      \expandafter\def\csname LTa\endcsname{\color{black}}%
      \expandafter\def\csname LT0\endcsname{\color[rgb]{1,0,0}}%
      \expandafter\def\csname LT1\endcsname{\color[rgb]{0,1,0}}%
      \expandafter\def\csname LT2\endcsname{\color[rgb]{0,0,1}}%
      \expandafter\def\csname LT3\endcsname{\color[rgb]{1,0,1}}%
      \expandafter\def\csname LT4\endcsname{\color[rgb]{0,1,1}}%
      \expandafter\def\csname LT5\endcsname{\color[rgb]{1,1,0}}%
      \expandafter\def\csname LT6\endcsname{\color[rgb]{0,0,0}}%
      \expandafter\def\csname LT7\endcsname{\color[rgb]{1,0.3,0}}%
      \expandafter\def\csname LT8\endcsname{\color[rgb]{0.5,0.5,0.5}}%
    \else
      % gray
      \def\colorrgb#1{\color{black}}%
      \def\colorgray#1{\color[gray]{#1}}%
      \expandafter\def\csname LTw\endcsname{\color{white}}%
      \expandafter\def\csname LTb\endcsname{\color{black}}%
      \expandafter\def\csname LTa\endcsname{\color{black}}%
      \expandafter\def\csname LT0\endcsname{\color{black}}%
      \expandafter\def\csname LT1\endcsname{\color{black}}%
      \expandafter\def\csname LT2\endcsname{\color{black}}%
      \expandafter\def\csname LT3\endcsname{\color{black}}%
      \expandafter\def\csname LT4\endcsname{\color{black}}%
      \expandafter\def\csname LT5\endcsname{\color{black}}%
      \expandafter\def\csname LT6\endcsname{\color{black}}%
      \expandafter\def\csname LT7\endcsname{\color{black}}%
      \expandafter\def\csname LT8\endcsname{\color{black}}%
    \fi
  \fi
    \setlength{\unitlength}{0.0500bp}%
    \ifx\gptboxheight\undefined%
      \newlength{\gptboxheight}%
      \newlength{\gptboxwidth}%
      \newsavebox{\gptboxtext}%
    \fi%
    \setlength{\fboxrule}{0.5pt}%
    \setlength{\fboxsep}{1pt}%
\begin{picture}(7200.00,5040.00)%
    \gplgaddtomacro\gplbacktext{%
      \csname LTb\endcsname%%
      \put(682,704){\makebox(0,0)[r]{\strut{}$0$}}%
      \put(682,1390){\makebox(0,0)[r]{\strut{}$5$}}%
      \put(682,2076){\makebox(0,0)[r]{\strut{}$10$}}%
      \put(682,2762){\makebox(0,0)[r]{\strut{}$15$}}%
      \put(682,3447){\makebox(0,0)[r]{\strut{}$20$}}%
      \put(682,4133){\makebox(0,0)[r]{\strut{}$25$}}%
      \put(682,4819){\makebox(0,0)[r]{\strut{}$30$}}%
      \put(1678,484){\makebox(0,0){\strut{}$1$}}%
      \put(3350,484){\makebox(0,0){\strut{}$10$}}%
      \put(5022,484){\makebox(0,0){\strut{}$100$}}%
      \put(6694,484){\makebox(0,0){\strut{}$1000$}}%
    }%
    \gplgaddtomacro\gplfronttext{%
      \csname LTb\endcsname%%
      \put(198,2761){\rotatebox{-270}{\makebox(0,0){ \large $\widetilde{u}^+$}}}%
      \put(3808,154){\makebox(0,0){\large $z^+$}}%
      \csname LTb\endcsname%%
      \put(5816,1317){\makebox(0,0)[r]{\strut{}LU model}}%
      \csname LTb\endcsname%%
      \put(5816,1097){\makebox(0,0)[r]{\strut{}DNS}}%
      \csname LTb\endcsname%%
      \put(5816,877){\makebox(0,0)[r]{\strut{}Classical Laws}}%
    }%
    \gplbacktext
    \put(0,0){\includegraphics{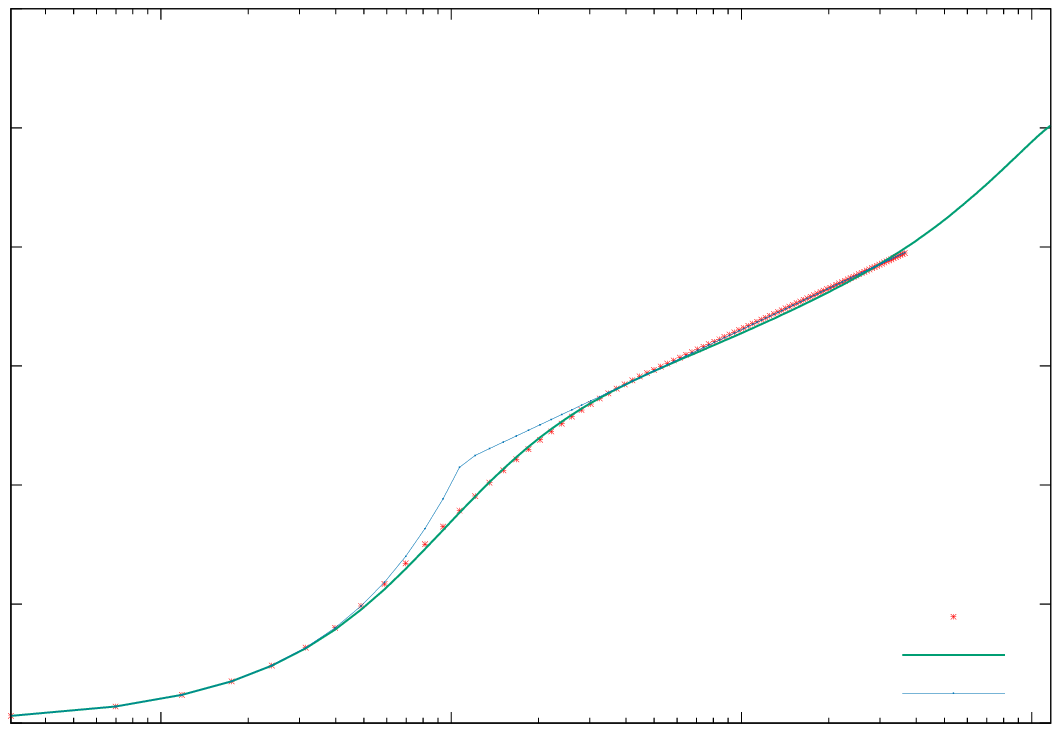}}%
    \gplfronttext
  \end{picture}%
\endgroup

%% file: UMean_TBL_1437.tex
% GNUPLOT: LaTeX picture with Postscript
\begingroup
  \makeatletter
  \providecommand\color[2][]{%
    \GenericError{(gnuplot) \space\space\space\@spaces}{%
      Package color not loaded in conjunction with
      terminal option `colourtext'%
    }{See the gnuplot documentation for explanation.%
    }{Either use 'blacktext' in gnuplot or load the package
      color.sty in LaTeX.}%
    \renewcommand\color[2][]{}%
  }%
  \providecommand\includegraphics[2][]{%
    \GenericError{(gnuplot) \space\space\space\@spaces}{%
      Package graphicx or graphics not loaded%
    }{See the gnuplot documentation for explanation.%
    }{The gnuplot epslatex terminal needs graphicx.sty or graphics.sty.}%
    \renewcommand\includegraphics[2][]{}%
  }%
  \providecommand\rotatebox[2]{#2}%
  \@ifundefined{ifGPcolor}{%
    \newif\ifGPcolor
    \GPcolorfalse
  }{}%
  \@ifundefined{ifGPblacktext}{%
    \newif\ifGPblacktext
    \GPblacktexttrue
  }{}%
  % define a \g@addto@macro without @ in the name:
  \let\gplgaddtomacro\g@addto@macro
  % define empty templates for all commands taking text:
  \gdef\gplbacktext{}%
  \gdef\gplfronttext{}%
  \makeatother
  \ifGPblacktext
    % no textcolor at all
    \def\colorrgb#1{}%
    \def\colorgray#1{}%
  \else
    % gray or color?
    \ifGPcolor
      \def\colorrgb#1{\color[rgb]{#1}}%
      \def\colorgray#1{\color[gray]{#1}}%
      \expandafter\def\csname LTw\endcsname{\color{white}}%
      \expandafter\def\csname LTb\endcsname{\color{black}}%
      \expandafter\def\csname LTa\endcsname{\color{black}}%
      \expandafter\def\csname LT0\endcsname{\color[rgb]{1,0,0}}%
      \expandafter\def\csname LT1\endcsname{\color[rgb]{0,1,0}}%
      \expandafter\def\csname LT2\endcsname{\color[rgb]{0,0,1}}%
      \expandafter\def\csname LT3\endcsname{\color[rgb]{1,0,1}}%
      \expandafter\def\csname LT4\endcsname{\color[rgb]{0,1,1}}%
      \expandafter\def\csname LT5\endcsname{\color[rgb]{1,1,0}}%
      \expandafter\def\csname LT6\endcsname{\color[rgb]{0,0,0}}%
      \expandafter\def\csname LT7\endcsname{\color[rgb]{1,0.3,0}}%
      \expandafter\def\csname LT8\endcsname{\color[rgb]{0.5,0.5,0.5}}%
    \else
      % gray
      \def\colorrgb#1{\color{black}}%
      \def\colorgray#1{\color[gray]{#1}}%
      \expandafter\def\csname LTw\endcsname{\color{white}}%
      \expandafter\def\csname LTb\endcsname{\color{black}}%
      \expandafter\def\csname LTa\endcsname{\color{black}}%
      \expandafter\def\csname LT0\endcsname{\color{black}}%
      \expandafter\def\csname LT1\endcsname{\color{black}}%
      \expandafter\def\csname LT2\endcsname{\color{black}}%
      \expandafter\def\csname LT3\endcsname{\color{black}}%
      \expandafter\def\csname LT4\endcsname{\color{black}}%
      \expandafter\def\csname LT5\endcsname{\color{black}}%
      \expandafter\def\csname LT6\endcsname{\color{black}}%
      \expandafter\def\csname LT7\endcsname{\color{black}}%
      \expandafter\def\csname LT8\endcsname{\color{black}}%
    \fi
  \fi
    \setlength{\unitlength}{0.0500bp}%
    \ifx\gptboxheight\undefined%
      \newlength{\gptboxheight}%
      \newlength{\gptboxwidth}%
      \newsavebox{\gptboxtext}%
    \fi%
    \setlength{\fboxrule}{0.5pt}%
    \setlength{\fboxsep}{1pt}%
\begin{picture}(7200.00,5040.00)%
    \gplgaddtomacro\gplbacktext{%
      \csname LTb\endcsname%%
      \put(682,704){\makebox(0,0)[r]{\strut{}$0$}}%
      \put(682,1527){\makebox(0,0)[r]{\strut{}$5$}}%
      \put(682,2350){\makebox(0,0)[r]{\strut{}$10$}}%
      \put(682,3173){\makebox(0,0)[r]{\strut{}$15$}}%
      \put(682,3996){\makebox(0,0)[r]{\strut{}$20$}}%
      \put(682,4819){\makebox(0,0)[r]{\strut{}$25$}}%
      \put(1686,484){\makebox(0,0){\strut{}$1$}}%
      \put(3358,484){\makebox(0,0){\strut{}$10$}}%
      \put(5030,484){\makebox(0,0){\strut{}$100$}}%
      \put(6702,484){\makebox(0,0){\strut{}$1000$}}%
    }%
    \gplgaddtomacro\gplfronttext{%
      \csname LTb\endcsname%%
      \put(198,2761){\rotatebox{-270}{\makebox(0,0){\large $\widetilde{u}^+$}}}%
      \put(3808,154){\makebox(0,0){\large $z^+$}}%
      \csname LTb\endcsname%%
      \put(5816,1317){\makebox(0,0)[r]{\strut{}LU model}}%
      \csname LTb\endcsname%%
      \put(5816,1097){\makebox(0,0)[r]{\strut{}DNS}}%
      \csname LTb\endcsname%%
      \put(5816,877){\makebox(0,0)[r]{\strut{}Classical Laws}}%
    }%
    \gplbacktext
    \put(0,0){\includegraphics{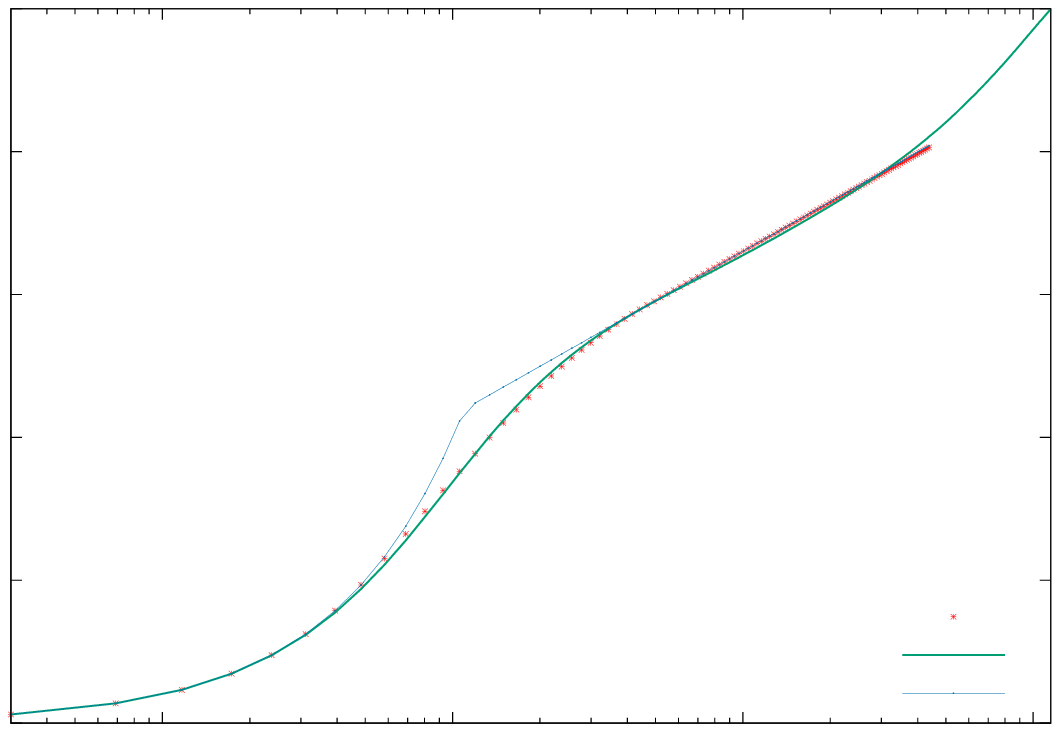}}%
    \gplfronttext
  \end{picture}%
\endgroup

%% file: UMean_TBL_1709.tex
% GNUPLOT: LaTeX picture with Postscript
\begingroup
  \makeatletter
  \providecommand\color[2][]{%
    \GenericError{(gnuplot) \space\space\space\@spaces}{%
      Package color not loaded in conjunction with
      terminal option `colourtext'%
    }{See the gnuplot documentation for explanation.%
    }{Either use 'blacktext' in gnuplot or load the package
      color.sty in LaTeX.}%
    \renewcommand\color[2][]{}%
  }%
  \providecommand\includegraphics[2][]{%
    \GenericError{(gnuplot) \space\space\space\@spaces}{%
      Package graphicx or graphics not loaded%
    }{See the gnuplot documentation for explanation.%
    }{The gnuplot epslatex terminal needs graphicx.sty or graphics.sty.}%
    \renewcommand\includegraphics[2][]{}%
  }%
  \providecommand\rotatebox[2]{#2}%
  \@ifundefined{ifGPcolor}{%
    \newif\ifGPcolor
    \GPcolorfalse
  }{}%
  \@ifundefined{ifGPblacktext}{%
    \newif\ifGPblacktext
    \GPblacktexttrue
  }{}%
  % define a \g@addto@macro without @ in the name:
  \let\gplgaddtomacro\g@addto@macro
  % define empty templates for all commands taking text:
  \gdef\gplbacktext{}%
  \gdef\gplfronttext{}%
  \makeatother
  \ifGPblacktext
    % no textcolor at all
    \def\colorrgb#1{}%
    \def\colorgray#1{}%
  \else
    % gray or color?
    \ifGPcolor
      \def\colorrgb#1{\color[rgb]{#1}}%
      \def\colorgray#1{\color[gray]{#1}}%
      \expandafter\def\csname LTw\endcsname{\color{white}}%
      \expandafter\def\csname LTb\endcsname{\color{black}}%
      \expandafter\def\csname LTa\endcsname{\color{black}}%
      \expandafter\def\csname LT0\endcsname{\color[rgb]{1,0,0}}%
      \expandafter\def\csname LT1\endcsname{\color[rgb]{0,1,0}}%
      \expandafter\def\csname LT2\endcsname{\color[rgb]{0,0,1}}%
      \expandafter\def\csname LT3\endcsname{\color[rgb]{1,0,1}}%
      \expandafter\def\csname LT4\endcsname{\color[rgb]{0,1,1}}%
      \expandafter\def\csname LT5\endcsname{\color[rgb]{1,1,0}}%
      \expandafter\def\csname LT6\endcsname{\color[rgb]{0,0,0}}%
      \expandafter\def\csname LT7\endcsname{\color[rgb]{1,0.3,0}}%
      \expandafter\def\csname LT8\endcsname{\color[rgb]{0.5,0.5,0.5}}%
    \else
      % gray
      \def\colorrgb#1{\color{black}}%
      \def\colorgray#1{\color[gray]{#1}}%
      \expandafter\def\csname LTw\endcsname{\color{white}}%
      \expandafter\def\csname LTb\endcsname{\color{black}}%
      \expandafter\def\csname LTa\endcsname{\color{black}}%
      \expandafter\def\csname LT0\endcsname{\color{black}}%
      \expandafter\def\csname LT1\endcsname{\color{black}}%
      \expandafter\def\csname LT2\endcsname{\color{black}}%
      \expandafter\def\csname LT3\endcsname{\color{black}}%
      \expandafter\def\csname LT4\endcsname{\color{black}}%
      \expandafter\def\csname LT5\endcsname{\color{black}}%
      \expandafter\def\csname LT6\endcsname{\color{black}}%
      \expandafter\def\csname LT7\endcsname{\color{black}}%
      \expandafter\def\csname LT8\endcsname{\color{black}}%
    \fi
  \fi
    \setlength{\unitlength}{0.0500bp}%
    \ifx\gptboxheight\undefined%
      \newlength{\gptboxheight}%
      \newlength{\gptboxwidth}%
      \newsavebox{\gptboxtext}%
    \fi%
    \setlength{\fboxrule}{0.5pt}%
    \setlength{\fboxsep}{1pt}%
\begin{picture}(7200.00,5040.00)%
    \gplgaddtomacro\gplbacktext{%
      \csname LTb\endcsname%%
      \put(682,704){\makebox(0,0)[r]{\strut{}$0$}}%
      \put(682,1390){\makebox(0,0)[r]{\strut{}$5$}}%
      \put(682,2076){\makebox(0,0)[r]{\strut{}$10$}}%
      \put(682,2762){\makebox(0,0)[r]{\strut{}$15$}}%
      \put(682,3447){\makebox(0,0)[r]{\strut{}$20$}}%
      \put(682,4133){\makebox(0,0)[r]{\strut{}$25$}}%
      \put(682,4819){\makebox(0,0)[r]{\strut{}$30$}}%
      \put(1673,484){\makebox(0,0){\strut{}$1$}}%
      \put(3288,484){\makebox(0,0){\strut{}$10$}}%
      \put(4902,484){\makebox(0,0){\strut{}$100$}}%
      \put(6516,484){\makebox(0,0){\strut{}$1000$}}%
    }%
    \gplgaddtomacro\gplfronttext{%
      \csname LTb\endcsname%%
      \put(198,2761){\rotatebox{-270}{\makebox(0,0){\large $\widetilde{u}^+$}}}%
      \put(3808,154){\makebox(0,0){\large $z^+$}}%
      \csname LTb\endcsname%%
      \put(5816,1317){\makebox(0,0)[r]{\strut{}LU model}}%
      \csname LTb\endcsname%%
      \put(5816,1097){\makebox(0,0)[r]{\strut{}DNS}}%
      \csname LTb\endcsname%%
      \put(5816,877){\makebox(0,0)[r]{\strut{}Classical Laws}}%
    }%
    \gplbacktext
    \put(0,0){\includegraphics{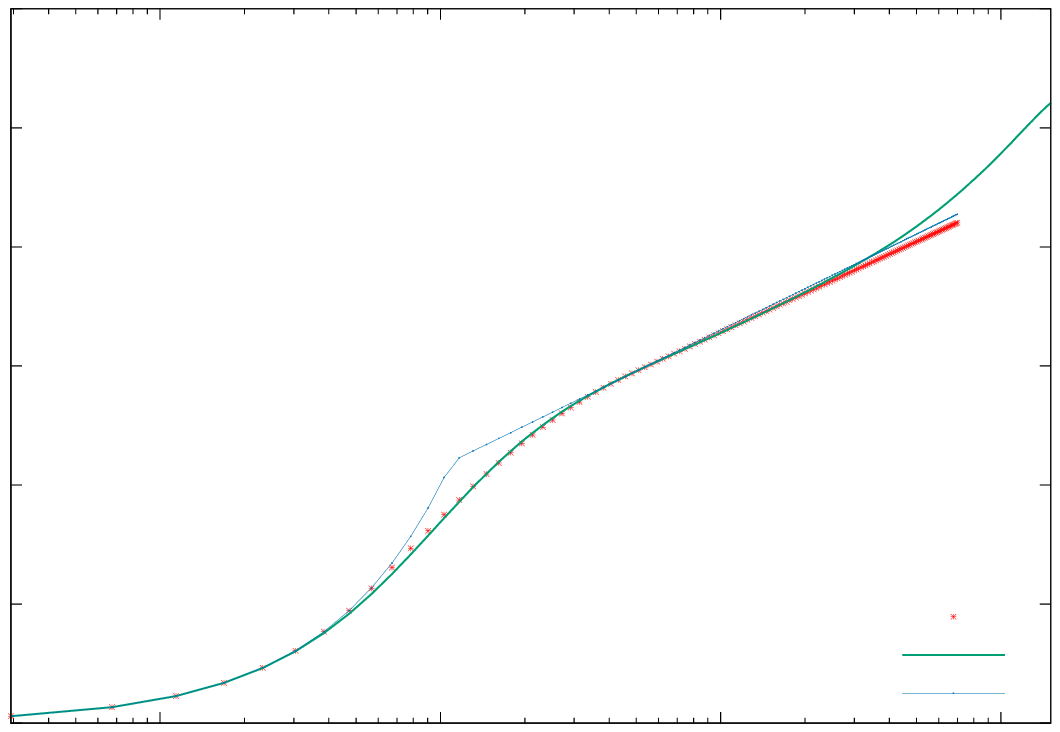}}%
    \gplfronttext
  \end{picture}%
\endgroup

%% file: UMean_TBL_1989.tex
% GNUPLOT: LaTeX picture with Postscript
\begingroup
  \makeatletter
  \providecommand\color[2][]{%
    \GenericError{(gnuplot) \space\space\space\@spaces}{%
      Package color not loaded in conjunction with
      terminal option `colourtext'%
    }{See the gnuplot documentation for explanation.%
    }{Either use 'blacktext' in gnuplot or load the package
      color.sty in LaTeX.}%
    \renewcommand\color[2][]{}%
  }%
  \providecommand\includegraphics[2][]{%
    \GenericError{(gnuplot) \space\space\space\@spaces}{%
      Package graphicx or graphics not loaded%
    }{See the gnuplot documentation for explanation.%
    }{The gnuplot epslatex terminal needs graphicx.sty or graphics.sty.}%
    \renewcommand\includegraphics[2][]{}%
  }%
  \providecommand\rotatebox[2]{#2}%
  \@ifundefined{ifGPcolor}{%
    \newif\ifGPcolor
    \GPcolorfalse
  }{}%
  \@ifundefined{ifGPblacktext}{%
    \newif\ifGPblacktext
    \GPblacktexttrue
  }{}%
  % define a \g@addto@macro without @ in the name:
  \let\gplgaddtomacro\g@addto@macro
  % define empty templates for all commands taking text:
  \gdef\gplbacktext{}%
  \gdef\gplfronttext{}%
  \makeatother
  \ifGPblacktext
    % no textcolor at all
    \def\colorrgb#1{}%
    \def\colorgray#1{}%
  \else
    % gray or color?
    \ifGPcolor
      \def\colorrgb#1{\color[rgb]{#1}}%
      \def\colorgray#1{\color[gray]{#1}}%
      \expandafter\def\csname LTw\endcsname{\color{white}}%
      \expandafter\def\csname LTb\endcsname{\color{black}}%
      \expandafter\def\csname LTa\endcsname{\color{black}}%
      \expandafter\def\csname LT0\endcsname{\color[rgb]{1,0,0}}%
      \expandafter\def\csname LT1\endcsname{\color[rgb]{0,1,0}}%
      \expandafter\def\csname LT2\endcsname{\color[rgb]{0,0,1}}%
      \expandafter\def\csname LT3\endcsname{\color[rgb]{1,0,1}}%
      \expandafter\def\csname LT4\endcsname{\color[rgb]{0,1,1}}%
      \expandafter\def\csname LT5\endcsname{\color[rgb]{1,1,0}}%
      \expandafter\def\csname LT6\endcsname{\color[rgb]{0,0,0}}%
      \expandafter\def\csname LT7\endcsname{\color[rgb]{1,0.3,0}}%
      \expandafter\def\csname LT8\endcsname{\color[rgb]{0.5,0.5,0.5}}%
    \else
      % gray
      \def\colorrgb#1{\color{black}}%
      \def\colorgray#1{\color[gray]{#1}}%
      \expandafter\def\csname LTw\endcsname{\color{white}}%
      \expandafter\def\csname LTb\endcsname{\color{black}}%
      \expandafter\def\csname LTa\endcsname{\color{black}}%
      \expandafter\def\csname LT0\endcsname{\color{black}}%
      \expandafter\def\csname LT1\endcsname{\color{black}}%
      \expandafter\def\csname LT2\endcsname{\color{black}}%
      \expandafter\def\csname LT3\endcsname{\color{black}}%
      \expandafter\def\csname LT4\endcsname{\color{black}}%
      \expandafter\def\csname LT5\endcsname{\color{black}}%
      \expandafter\def\csname LT6\endcsname{\color{black}}%
      \expandafter\def\csname LT7\endcsname{\color{black}}%
      \expandafter\def\csname LT8\endcsname{\color{black}}%
    \fi
  \fi
    \setlength{\unitlength}{0.0500bp}%
    \ifx\gptboxheight\undefined%
      \newlength{\gptboxheight}%
      \newlength{\gptboxwidth}%
      \newsavebox{\gptboxtext}%
    \fi%
    \setlength{\fboxrule}{0.5pt}%
    \setlength{\fboxsep}{1pt}%
\begin{picture}(7200.00,5040.00)%
    \gplgaddtomacro\gplbacktext{%
      \csname LTb\endcsname%%
      \put(682,704){\makebox(0,0)[r]{\strut{}$0$}}%
      \put(682,1390){\makebox(0,0)[r]{\strut{}$5$}}%
      \put(682,2076){\makebox(0,0)[r]{\strut{}$10$}}%
      \put(682,2762){\makebox(0,0)[r]{\strut{}$15$}}%
      \put(682,3447){\makebox(0,0)[r]{\strut{}$20$}}%
      \put(682,4133){\makebox(0,0)[r]{\strut{}$25$}}%
      \put(682,4819){\makebox(0,0)[r]{\strut{}$30$}}%
      \put(1678,484){\makebox(0,0){\strut{}$1$}}%
      \put(3280,484){\makebox(0,0){\strut{}$10$}}%
      \put(4882,484){\makebox(0,0){\strut{}$100$}}%
      \put(6485,484){\makebox(0,0){\strut{}$1000$}}%
    }%
    \gplgaddtomacro\gplfronttext{%
      \csname LTb\endcsname%%
      \put(198,2761){\rotatebox{-270}{\makebox(0,0){\large $\widetilde{u}^+$}}}%
      \put(3808,154){\makebox(0,0){\large $z^+$}}%
      \csname LTb\endcsname%%
      \put(5816,1317){\makebox(0,0)[r]{\strut{}LU model}}%
      \csname LTb\endcsname%%
      \put(5816,1097){\makebox(0,0)[r]{\strut{}DNS}}%
      \csname LTb\endcsname%%
      \put(5816,877){\makebox(0,0)[r]{\strut{}Classical Laws}}%
    }%
    \gplbacktext
    \put(0,0){\includegraphics{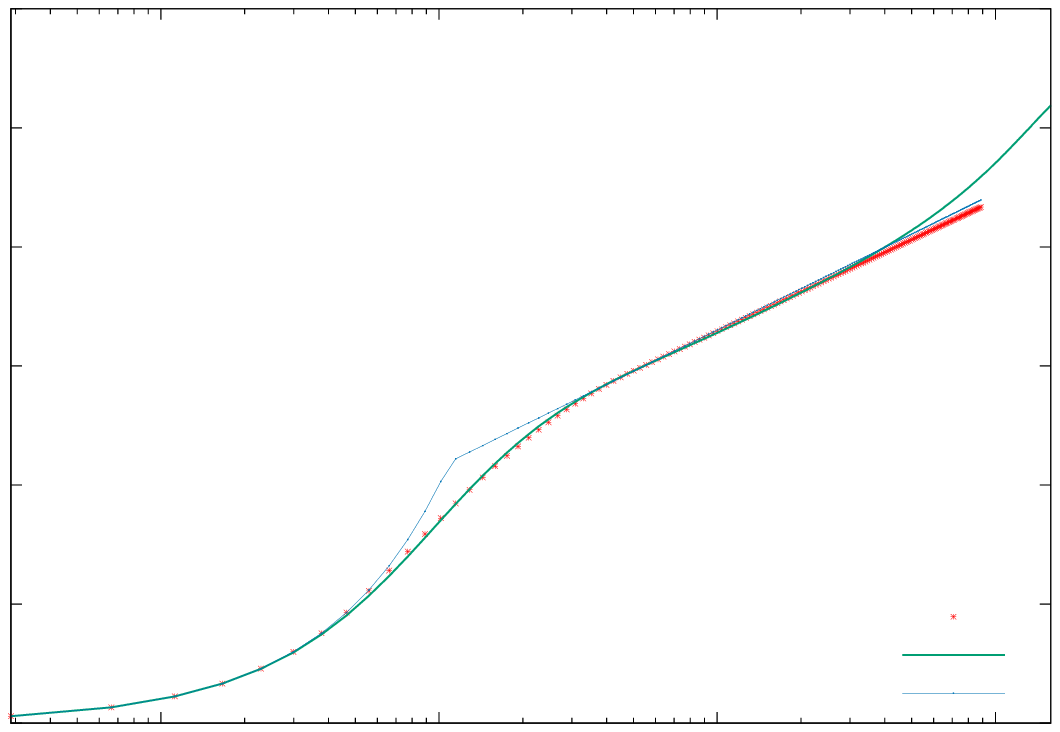}}%
    \gplfronttext
  \end{picture}%
\endgroup

%% file: UMean_Pipe_180.tex
% GNUPLOT: LaTeX picture with Postscript
\begingroup
  \makeatletter
  \providecommand\color[2][]{%
    \GenericError{(gnuplot) \space\space\space\@spaces}{%
      Package color not loaded in conjunction with
      terminal option `colourtext'%
    }{See the gnuplot documentation for explanation.%
    }{Either use 'blacktext' in gnuplot or load the package
      color.sty in LaTeX.}%
    \renewcommand\color[2][]{}%
  }%
  \providecommand\includegraphics[2][]{%
    \GenericError{(gnuplot) \space\space\space\@spaces}{%
      Package graphicx or graphics not loaded%
    }{See the gnuplot documentation for explanation.%
    }{The gnuplot epslatex terminal needs graphicx.sty or graphics.sty.}%
    \renewcommand\includegraphics[2][]{}%
  }%
  \providecommand\rotatebox[2]{#2}%
  \@ifundefined{ifGPcolor}{%
    \newif\ifGPcolor
    \GPcolorfalse
  }{}%
  \@ifundefined{ifGPblacktext}{%
    \newif\ifGPblacktext
    \GPblacktexttrue
  }{}%
  % define a \g@addto@macro without @ in the name:
  \let\gplgaddtomacro\g@addto@macro
  % define empty templates for all commands taking text:
  \gdef\gplbacktext{}%
  \gdef\gplfronttext{}%
  \makeatother
  \ifGPblacktext
    % no textcolor at all
    \def\colorrgb#1{}%
    \def\colorgray#1{}%
  \else
    % gray or color?
    \ifGPcolor
      \def\colorrgb#1{\color[rgb]{#1}}%
      \def\colorgray#1{\color[gray]{#1}}%
      \expandafter\def\csname LTw\endcsname{\color{white}}%
      \expandafter\def\csname LTb\endcsname{\color{black}}%
      \expandafter\def\csname LTa\endcsname{\color{black}}%
      \expandafter\def\csname LT0\endcsname{\color[rgb]{1,0,0}}%
      \expandafter\def\csname LT1\endcsname{\color[rgb]{0,1,0}}%
      \expandafter\def\csname LT2\endcsname{\color[rgb]{0,0,1}}%
      \expandafter\def\csname LT3\endcsname{\color[rgb]{1,0,1}}%
      \expandafter\def\csname LT4\endcsname{\color[rgb]{0,1,1}}%
      \expandafter\def\csname LT5\endcsname{\color[rgb]{1,1,0}}%
      \expandafter\def\csname LT6\endcsname{\color[rgb]{0,0,0}}%
      \expandafter\def\csname LT7\endcsname{\color[rgb]{1,0.3,0}}%
      \expandafter\def\csname LT8\endcsname{\color[rgb]{0.5,0.5,0.5}}%
    \else
      % gray
      \def\colorrgb#1{\color{black}}%
      \def\colorgray#1{\color[gray]{#1}}%
      \expandafter\def\csname LTw\endcsname{\color{white}}%
      \expandafter\def\csname LTb\endcsname{\color{black}}%
      \expandafter\def\csname LTa\endcsname{\color{black}}%
      \expandafter\def\csname LT0\endcsname{\color{black}}%
      \expandafter\def\csname LT1\endcsname{\color{black}}%
      \expandafter\def\csname LT2\endcsname{\color{black}}%
      \expandafter\def\csname LT3\endcsname{\color{black}}%
      \expandafter\def\csname LT4\endcsname{\color{black}}%
      \expandafter\def\csname LT5\endcsname{\color{black}}%
      \expandafter\def\csname LT6\endcsname{\color{black}}%
      \expandafter\def\csname LT7\endcsname{\color{black}}%
      \expandafter\def\csname LT8\endcsname{\color{black}}%
    \fi
  \fi
    \setlength{\unitlength}{0.0500bp}%
    \ifx\gptboxheight\undefined%
      \newlength{\gptboxheight}%
      \newlength{\gptboxwidth}%
      \newsavebox{\gptboxtext}%
    \fi%
    \setlength{\fboxrule}{0.5pt}%
    \setlength{\fboxsep}{1pt}%
\begin{picture}(7200.00,5040.00)%
    \gplgaddtomacro\gplbacktext{%
      \csname LTb\endcsname%%
      \put(682,704){\makebox(0,0)[r]{\strut{}$0$}}%
      \put(682,1116){\makebox(0,0)[r]{\strut{}$2$}}%
      \put(682,1527){\makebox(0,0)[r]{\strut{}$4$}}%
      \put(682,1939){\makebox(0,0)[r]{\strut{}$6$}}%
      \put(682,2350){\makebox(0,0)[r]{\strut{}$8$}}%
      \put(682,2762){\makebox(0,0)[r]{\strut{}$10$}}%
      \put(682,3173){\makebox(0,0)[r]{\strut{}$12$}}%
      \put(682,3585){\makebox(0,0)[r]{\strut{}$14$}}%
      \put(682,3996){\makebox(0,0)[r]{\strut{}$16$}}%
      \put(682,4408){\makebox(0,0)[r]{\strut{}$18$}}%
      \put(682,4819){\makebox(0,0)[r]{\strut{}$20$}}%
      \put(2446,484){\makebox(0,0){\strut{}$1$}}%
      \put(4376,484){\makebox(0,0){\strut{}$10$}}%
      \put(6306,484){\makebox(0,0){\strut{}$100$}}%
    }%
    \gplgaddtomacro\gplfronttext{%
      \csname LTb\endcsname%%
      \put(198,2761){\rotatebox{-270}{\makebox(0,0){\large $\widetilde{u}^+$}}}%
      \put(3808,154){\makebox(0,0){\large $z^+$}}%
      \csname LTb\endcsname%%
      \put(5816,1317){\makebox(0,0)[r]{\strut{}LU model}}%
      \csname LTb\endcsname%%
      \put(5816,1097){\makebox(0,0)[r]{\strut{}DNS}}%
      \csname LTb\endcsname%%
      \put(5816,877){\makebox(0,0)[r]{\strut{}Classical Laws}}%
    }%
    \gplbacktext
    \put(0,0){\includegraphics{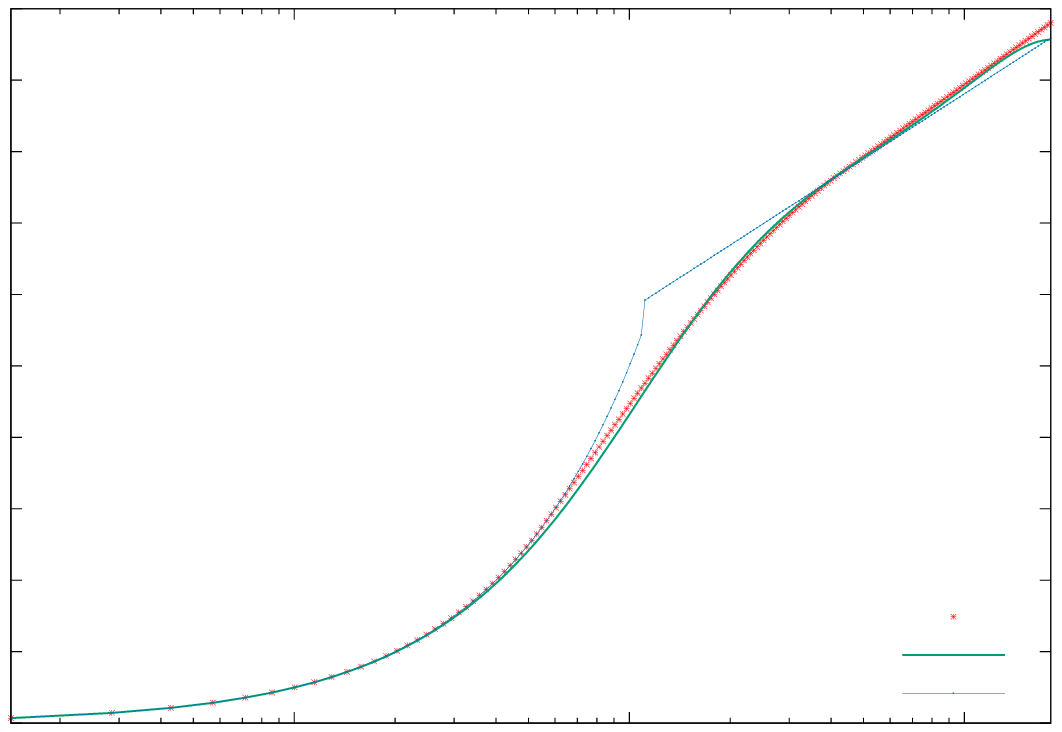}}%
    \gplfronttext
  \end{picture}%
\endgroup

%% file: UMean_Pipe_360.tex
% GNUPLOT: LaTeX picture with Postscript
\begingroup
  \makeatletter
  \providecommand\color[2][]{%
    \GenericError{(gnuplot) \space\space\space\@spaces}{%
      Package color not loaded in conjunction with
      terminal option `colourtext'%
    }{See the gnuplot documentation for explanation.%
    }{Either use 'blacktext' in gnuplot or load the package
      color.sty in LaTeX.}%
    \renewcommand\color[2][]{}%
  }%
  \providecommand\includegraphics[2][]{%
    \GenericError{(gnuplot) \space\space\space\@spaces}{%
      Package graphicx or graphics not loaded%
    }{See the gnuplot documentation for explanation.%
    }{The gnuplot epslatex terminal needs graphicx.sty or graphics.sty.}%
    \renewcommand\includegraphics[2][]{}%
  }%
  \providecommand\rotatebox[2]{#2}%
  \@ifundefined{ifGPcolor}{%
    \newif\ifGPcolor
    \GPcolorfalse
  }{}%
  \@ifundefined{ifGPblacktext}{%
    \newif\ifGPblacktext
    \GPblacktexttrue
  }{}%
  % define a \g@addto@macro without @ in the name:
  \let\gplgaddtomacro\g@addto@macro
  % define empty templates for all commands taking text:
  \gdef\gplbacktext{}%
  \gdef\gplfronttext{}%
  \makeatother
  \ifGPblacktext
    % no textcolor at all
    \def\colorrgb#1{}%
    \def\colorgray#1{}%
  \else
    % gray or color?
    \ifGPcolor
      \def\colorrgb#1{\color[rgb]{#1}}%
      \def\colorgray#1{\color[gray]{#1}}%
      \expandafter\def\csname LTw\endcsname{\color{white}}%
      \expandafter\def\csname LTb\endcsname{\color{black}}%
      \expandafter\def\csname LTa\endcsname{\color{black}}%
      \expandafter\def\csname LT0\endcsname{\color[rgb]{1,0,0}}%
      \expandafter\def\csname LT1\endcsname{\color[rgb]{0,1,0}}%
      \expandafter\def\csname LT2\endcsname{\color[rgb]{0,0,1}}%
      \expandafter\def\csname LT3\endcsname{\color[rgb]{1,0,1}}%
      \expandafter\def\csname LT4\endcsname{\color[rgb]{0,1,1}}%
      \expandafter\def\csname LT5\endcsname{\color[rgb]{1,1,0}}%
      \expandafter\def\csname LT6\endcsname{\color[rgb]{0,0,0}}%
      \expandafter\def\csname LT7\endcsname{\color[rgb]{1,0.3,0}}%
      \expandafter\def\csname LT8\endcsname{\color[rgb]{0.5,0.5,0.5}}%
    \else
      % gray
      \def\colorrgb#1{\color{black}}%
      \def\colorgray#1{\color[gray]{#1}}%
      \expandafter\def\csname LTw\endcsname{\color{white}}%
      \expandafter\def\csname LTb\endcsname{\color{black}}%
      \expandafter\def\csname LTa\endcsname{\color{black}}%
      \expandafter\def\csname LT0\endcsname{\color{black}}%
      \expandafter\def\csname LT1\endcsname{\color{black}}%
      \expandafter\def\csname LT2\endcsname{\color{black}}%
      \expandafter\def\csname LT3\endcsname{\color{black}}%
      \expandafter\def\csname LT4\endcsname{\color{black}}%
      \expandafter\def\csname LT5\endcsname{\color{black}}%
      \expandafter\def\csname LT6\endcsname{\color{black}}%
      \expandafter\def\csname LT7\endcsname{\color{black}}%
      \expandafter\def\csname LT8\endcsname{\color{black}}%
    \fi
  \fi
    \setlength{\unitlength}{0.0500bp}%
    \ifx\gptboxheight\undefined%
      \newlength{\gptboxheight}%
      \newlength{\gptboxwidth}%
      \newsavebox{\gptboxtext}%
    \fi%
    \setlength{\fboxrule}{0.5pt}%
    \setlength{\fboxsep}{1pt}%
\begin{picture}(7200.00,5040.00)%
    \gplgaddtomacro\gplbacktext{%
      \csname LTb\endcsname%%
      \put(682,704){\makebox(0,0)[r]{\strut{}$0$}}%
      \put(682,1527){\makebox(0,0)[r]{\strut{}$5$}}%
      \put(682,2350){\makebox(0,0)[r]{\strut{}$10$}}%
      \put(682,3173){\makebox(0,0)[r]{\strut{}$15$}}%
      \put(682,3996){\makebox(0,0)[r]{\strut{}$20$}}%
      \put(682,4819){\makebox(0,0)[r]{\strut{}$25$}}%
      \put(2212,484){\makebox(0,0){\strut{}$1$}}%
      \put(4008,484){\makebox(0,0){\strut{}$10$}}%
      \put(5804,484){\makebox(0,0){\strut{}$100$}}%
    }%
    \gplgaddtomacro\gplfronttext{%
      \csname LTb\endcsname%%
      \put(198,2761){\rotatebox{-270}{\makebox(0,0){\large $\widetilde{u}^+$}}}%
      \put(3808,154){\makebox(0,0){\large $z^+$}}%
      \csname LTb\endcsname%%
      \put(5816,1317){\makebox(0,0)[r]{\strut{}LU model}}%
      \csname LTb\endcsname%%
      \put(5816,1097){\makebox(0,0)[r]{\strut{}DNS}}%
      \csname LTb\endcsname%%
      \put(5816,877){\makebox(0,0)[r]{\strut{}Classical Laws}}%
    }%
    \gplbacktext
    \put(0,0){\includegraphics{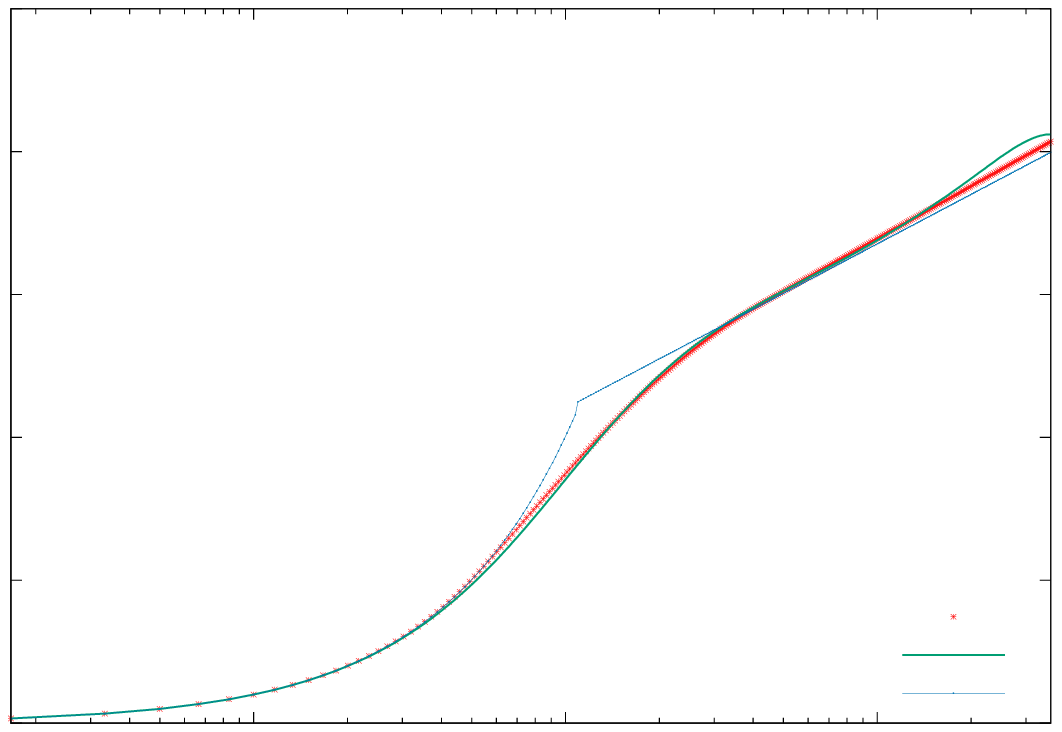}}%
    \gplfronttext
  \end{picture}%
\endgroup

%% file: UMean_Pipe_550.tex
% GNUPLOT: LaTeX picture with Postscript
\begingroup
  \makeatletter
  \providecommand\color[2][]{%
    \GenericError{(gnuplot) \space\space\space\@spaces}{%
      Package color not loaded in conjunction with
      terminal option `colourtext'%
    }{See the gnuplot documentation for explanation.%
    }{Either use 'blacktext' in gnuplot or load the package
      color.sty in LaTeX.}%
    \renewcommand\color[2][]{}%
  }%
  \providecommand\includegraphics[2][]{%
    \GenericError{(gnuplot) \space\space\space\@spaces}{%
      Package graphicx or graphics not loaded%
    }{See the gnuplot documentation for explanation.%
    }{The gnuplot epslatex terminal needs graphicx.sty or graphics.sty.}%
    \renewcommand\includegraphics[2][]{}%
  }%
  \providecommand\rotatebox[2]{#2}%
  \@ifundefined{ifGPcolor}{%
    \newif\ifGPcolor
    \GPcolorfalse
  }{}%
  \@ifundefined{ifGPblacktext}{%
    \newif\ifGPblacktext
    \GPblacktexttrue
  }{}%
  % define a \g@addto@macro without @ in the name:
  \let\gplgaddtomacro\g@addto@macro
  % define empty templates for all commands taking text:
  \gdef\gplbacktext{}%
  \gdef\gplfronttext{}%
  \makeatother
  \ifGPblacktext
    % no textcolor at all
    \def\colorrgb#1{}%
    \def\colorgray#1{}%
  \else
    % gray or color?
    \ifGPcolor
      \def\colorrgb#1{\color[rgb]{#1}}%
      \def\colorgray#1{\color[gray]{#1}}%
      \expandafter\def\csname LTw\endcsname{\color{white}}%
      \expandafter\def\csname LTb\endcsname{\color{black}}%
      \expandafter\def\csname LTa\endcsname{\color{black}}%
      \expandafter\def\csname LT0\endcsname{\color[rgb]{1,0,0}}%
      \expandafter\def\csname LT1\endcsname{\color[rgb]{0,1,0}}%
      \expandafter\def\csname LT2\endcsname{\color[rgb]{0,0,1}}%
      \expandafter\def\csname LT3\endcsname{\color[rgb]{1,0,1}}%
      \expandafter\def\csname LT4\endcsname{\color[rgb]{0,1,1}}%
      \expandafter\def\csname LT5\endcsname{\color[rgb]{1,1,0}}%
      \expandafter\def\csname LT6\endcsname{\color[rgb]{0,0,0}}%
      \expandafter\def\csname LT7\endcsname{\color[rgb]{1,0.3,0}}%
      \expandafter\def\csname LT8\endcsname{\color[rgb]{0.5,0.5,0.5}}%
    \else
      % gray
      \def\colorrgb#1{\color{black}}%
      \def\colorgray#1{\color[gray]{#1}}%
      \expandafter\def\csname LTw\endcsname{\color{white}}%
      \expandafter\def\csname LTb\endcsname{\color{black}}%
      \expandafter\def\csname LTa\endcsname{\color{black}}%
      \expandafter\def\csname LT0\endcsname{\color{black}}%
      \expandafter\def\csname LT1\endcsname{\color{black}}%
      \expandafter\def\csname LT2\endcsname{\color{black}}%
      \expandafter\def\csname LT3\endcsname{\color{black}}%
      \expandafter\def\csname LT4\endcsname{\color{black}}%
      \expandafter\def\csname LT5\endcsname{\color{black}}%
      \expandafter\def\csname LT6\endcsname{\color{black}}%
      \expandafter\def\csname LT7\endcsname{\color{black}}%
      \expandafter\def\csname LT8\endcsname{\color{black}}%
    \fi
  \fi
    \setlength{\unitlength}{0.0500bp}%
    \ifx\gptboxheight\undefined%
      \newlength{\gptboxheight}%
      \newlength{\gptboxwidth}%
      \newsavebox{\gptboxtext}%
    \fi%
    \setlength{\fboxrule}{0.5pt}%
    \setlength{\fboxsep}{1pt}%
\begin{picture}(7200.00,5040.00)%
    \gplgaddtomacro\gplbacktext{%
      \csname LTb\endcsname%%
      \put(682,704){\makebox(0,0)[r]{\strut{}$0$}}%
      \put(682,1527){\makebox(0,0)[r]{\strut{}$5$}}%
      \put(682,2350){\makebox(0,0)[r]{\strut{}$10$}}%
      \put(682,3173){\makebox(0,0)[r]{\strut{}$15$}}%
      \put(682,3996){\makebox(0,0)[r]{\strut{}$20$}}%
      \put(682,4819){\makebox(0,0)[r]{\strut{}$25$}}%
      \put(2196,484){\makebox(0,0){\strut{}$1$}}%
      \put(3877,484){\makebox(0,0){\strut{}$10$}}%
      \put(5558,484){\makebox(0,0){\strut{}$100$}}%
    }%
    \gplgaddtomacro\gplfronttext{%
      \csname LTb\endcsname%%
      \put(198,2761){\rotatebox{-270}{\makebox(0,0){\large $\widetilde{u}^+$}}}%
      \put(3808,154){\makebox(0,0){\large $z^+$}}%
      \csname LTb\endcsname%%
      \put(5816,1317){\makebox(0,0)[r]{\strut{}LU model}}%
      \csname LTb\endcsname%%
      \put(5816,1097){\makebox(0,0)[r]{\strut{}DNS}}%
      \csname LTb\endcsname%%
      \put(5816,877){\makebox(0,0)[r]{\strut{}Classical Laws}}%
    }%
    \gplbacktext
    \put(0,0){\includegraphics{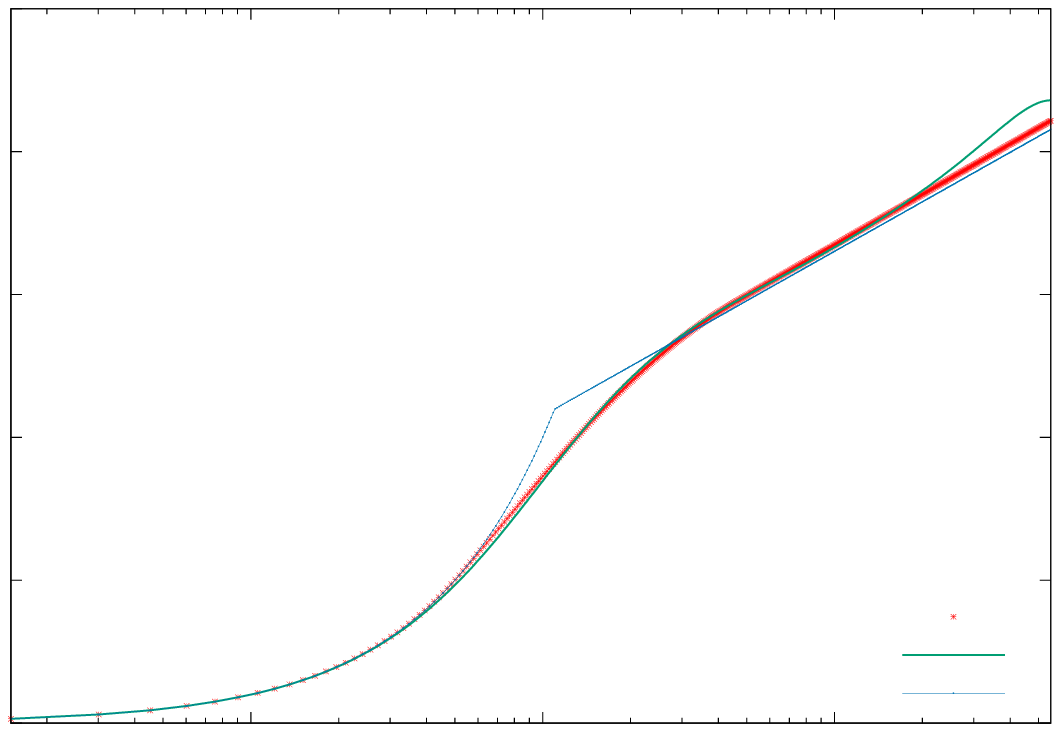}}%
    \gplfronttext
  \end{picture}%
\endgroup

%% file: UMean_Pipe_1000.tex
% GNUPLOT: LaTeX picture with Postscript
\begingroup
  \makeatletter
  \providecommand\color[2][]{%
    \GenericError{(gnuplot) \space\space\space\@spaces}{%
      Package color not loaded in conjunction with
      terminal option `colourtext'%
    }{See the gnuplot documentation for explanation.%
    }{Either use 'blacktext' in gnuplot or load the package
      color.sty in LaTeX.}%
    \renewcommand\color[2][]{}%
  }%
  \providecommand\includegraphics[2][]{%
    \GenericError{(gnuplot) \space\space\space\@spaces}{%
      Package graphicx or graphics not loaded%
    }{See the gnuplot documentation for explanation.%
    }{The gnuplot epslatex terminal needs graphicx.sty or graphics.sty.}%
    \renewcommand\includegraphics[2][]{}%
  }%
  \providecommand\rotatebox[2]{#2}%
  \@ifundefined{ifGPcolor}{%
    \newif\ifGPcolor
    \GPcolorfalse
  }{}%
  \@ifundefined{ifGPblacktext}{%
    \newif\ifGPblacktext
    \GPblacktexttrue
  }{}%
  % define a \g@addto@macro without @ in the name:
  \let\gplgaddtomacro\g@addto@macro
  % define empty templates for all commands taking text:
  \gdef\gplbacktext{}%
  \gdef\gplfronttext{}%
  \makeatother
  \ifGPblacktext
    % no textcolor at all
    \def\colorrgb#1{}%
    \def\colorgray#1{}%
  \else
    % gray or color?
    \ifGPcolor
      \def\colorrgb#1{\color[rgb]{#1}}%
      \def\colorgray#1{\color[gray]{#1}}%
      \expandafter\def\csname LTw\endcsname{\color{white}}%
      \expandafter\def\csname LTb\endcsname{\color{black}}%
      \expandafter\def\csname LTa\endcsname{\color{black}}%
      \expandafter\def\csname LT0\endcsname{\color[rgb]{1,0,0}}%
      \expandafter\def\csname LT1\endcsname{\color[rgb]{0,1,0}}%
      \expandafter\def\csname LT2\endcsname{\color[rgb]{0,0,1}}%
      \expandafter\def\csname LT3\endcsname{\color[rgb]{1,0,1}}%
      \expandafter\def\csname LT4\endcsname{\color[rgb]{0,1,1}}%
      \expandafter\def\csname LT5\endcsname{\color[rgb]{1,1,0}}%
      \expandafter\def\csname LT6\endcsname{\color[rgb]{0,0,0}}%
      \expandafter\def\csname LT7\endcsname{\color[rgb]{1,0.3,0}}%
      \expandafter\def\csname LT8\endcsname{\color[rgb]{0.5,0.5,0.5}}%
    \else
      % gray
      \def\colorrgb#1{\color{black}}%
      \def\colorgray#1{\color[gray]{#1}}%
      \expandafter\def\csname LTw\endcsname{\color{white}}%
      \expandafter\def\csname LTb\endcsname{\color{black}}%
      \expandafter\def\csname LTa\endcsname{\color{black}}%
      \expandafter\def\csname LT0\endcsname{\color{black}}%
      \expandafter\def\csname LT1\endcsname{\color{black}}%
      \expandafter\def\csname LT2\endcsname{\color{black}}%
      \expandafter\def\csname LT3\endcsname{\color{black}}%
      \expandafter\def\csname LT4\endcsname{\color{black}}%
      \expandafter\def\csname LT5\endcsname{\color{black}}%
      \expandafter\def\csname LT6\endcsname{\color{black}}%
      \expandafter\def\csname LT7\endcsname{\color{black}}%
      \expandafter\def\csname LT8\endcsname{\color{black}}%
    \fi
  \fi
    \setlength{\unitlength}{0.0500bp}%
    \ifx\gptboxheight\undefined%
      \newlength{\gptboxheight}%
      \newlength{\gptboxwidth}%
      \newsavebox{\gptboxtext}%
    \fi%
    \setlength{\fboxrule}{0.5pt}%
    \setlength{\fboxsep}{1pt}%
\begin{picture}(7200.00,5040.00)%
    \gplgaddtomacro\gplbacktext{%
      \csname LTb\endcsname%%
      \put(682,704){\makebox(0,0)[r]{\strut{}$0$}}%
      \put(682,1527){\makebox(0,0)[r]{\strut{}$5$}}%
      \put(682,2350){\makebox(0,0)[r]{\strut{}$10$}}%
      \put(682,3173){\makebox(0,0)[r]{\strut{}$15$}}%
      \put(682,3996){\makebox(0,0)[r]{\strut{}$20$}}%
      \put(682,4819){\makebox(0,0)[r]{\strut{}$25$}}%
      \put(2099,484){\makebox(0,0){\strut{}$1$}}%
      \put(3667,484){\makebox(0,0){\strut{}$10$}}%
      \put(5235,484){\makebox(0,0){\strut{}$100$}}%
      \put(6803,484){\makebox(0,0){\strut{}$1000$}}%
    }%
    \gplgaddtomacro\gplfronttext{%
      \csname LTb\endcsname%%
      \put(198,2761){\rotatebox{-270}{\makebox(0,0){\large $\widetilde{u}^+$}}}%
      \put(3808,154){\makebox(0,0){\large $z^+$}}%
      \csname LTb\endcsname%%
      \put(5816,1317){\makebox(0,0)[r]{\strut{}LU model}}%
      \csname LTb\endcsname%%
      \put(5816,1097){\makebox(0,0)[r]{\strut{}DNS}}%
      \csname LTb\endcsname%%
      \put(5816,877){\makebox(0,0)[r]{\strut{}Classical Laws}}%
    }%
    \gplbacktext
    \put(0,0){\includegraphics{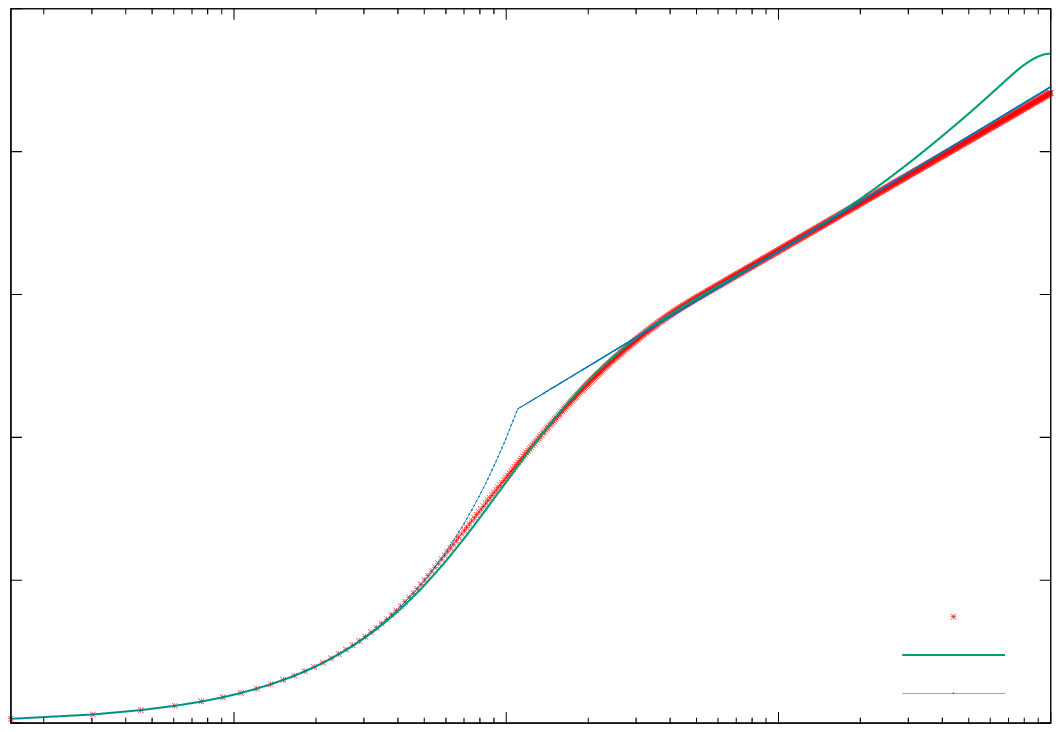}}%
    \gplfronttext
  \end{picture}%
\endgroup

%% file: UMean_5200.tex
% GNUPLOT: LaTeX picture with Postscript
\begingroup
  \makeatletter
  \providecommand\color[2][]{%
    \GenericError{(gnuplot) \space\space\space\@spaces}{%
      Package color not loaded in conjunction with
      terminal option `colourtext'%
    }{See the gnuplot documentation for explanation.%
    }{Either use 'blacktext' in gnuplot or load the package
      color.sty in LaTeX.}%
    \renewcommand\color[2][]{}%
  }%
  \providecommand\includegraphics[2][]{%
    \GenericError{(gnuplot) \space\space\space\@spaces}{%
      Package graphicx or graphics not loaded%
    }{See the gnuplot documentation for explanation.%
    }{The gnuplot epslatex terminal needs graphicx.sty or graphics.sty.}%
    \renewcommand\includegraphics[2][]{}%
  }%
  \providecommand\rotatebox[2]{#2}%
  \@ifundefined{ifGPcolor}{%
    \newif\ifGPcolor
    \GPcolorfalse
  }{}%
  \@ifundefined{ifGPblacktext}{%
    \newif\ifGPblacktext
    \GPblacktexttrue
  }{}%
  % define a \g@addto@macro without @ in the name:
  \let\gplgaddtomacro\g@addto@macro
  % define empty templates for all commands taking text:
  \gdef\gplbacktext{}%
  \gdef\gplfronttext{}%
  \makeatother
  \ifGPblacktext
    % no textcolor at all
    \def\colorrgb#1{}%
    \def\colorgray#1{}%
  \else
    % gray or color?
    \ifGPcolor
      \def\colorrgb#1{\color[rgb]{#1}}%
      \def\colorgray#1{\color[gray]{#1}}%
      \expandafter\def\csname LTw\endcsname{\color{white}}%
      \expandafter\def\csname LTb\endcsname{\color{black}}%
      \expandafter\def\csname LTa\endcsname{\color{black}}%
      \expandafter\def\csname LT0\endcsname{\color[rgb]{1,0,0}}%
      \expandafter\def\csname LT1\endcsname{\color[rgb]{0,1,0}}%
      \expandafter\def\csname LT2\endcsname{\color[rgb]{0,0,1}}%
      \expandafter\def\csname LT3\endcsname{\color[rgb]{1,0,1}}%
      \expandafter\def\csname LT4\endcsname{\color[rgb]{0,1,1}}%
      \expandafter\def\csname LT5\endcsname{\color[rgb]{1,1,0}}%
      \expandafter\def\csname LT6\endcsname{\color[rgb]{0,0,0}}%
      \expandafter\def\csname LT7\endcsname{\color[rgb]{1,0.3,0}}%
      \expandafter\def\csname LT8\endcsname{\color[rgb]{0.5,0.5,0.5}}%
    \else
      % gray
      \def\colorrgb#1{\color{black}}%
      \def\colorgray#1{\color[gray]{#1}}%
      \expandafter\def\csname LTw\endcsname{\color{white}}%
      \expandafter\def\csname LTb\endcsname{\color{black}}%
      \expandafter\def\csname LTa\endcsname{\color{black}}%
      \expandafter\def\csname LT0\endcsname{\color{black}}%
      \expandafter\def\csname LT1\endcsname{\color{black}}%
      \expandafter\def\csname LT2\endcsname{\color{black}}%
      \expandafter\def\csname LT3\endcsname{\color{black}}%
      \expandafter\def\csname LT4\endcsname{\color{black}}%
      \expandafter\def\csname LT5\endcsname{\color{black}}%
      \expandafter\def\csname LT6\endcsname{\color{black}}%
      \expandafter\def\csname LT7\endcsname{\color{black}}%
      \expandafter\def\csname LT8\endcsname{\color{black}}%
    \fi
  \fi
    \setlength{\unitlength}{0.0500bp}%
    \ifx\gptboxheight\undefined%
      \newlength{\gptboxheight}%
      \newlength{\gptboxwidth}%
      \newsavebox{\gptboxtext}%
    \fi%
    \setlength{\fboxrule}{0.5pt}%
    \setlength{\fboxsep}{1pt}%
\begin{picture}(7200.00,5040.00)%
    \gplgaddtomacro\gplbacktext{%
      \csname LTb\endcsname%
      \put(682,704){\makebox(0,0)[r]{\strut{}$0$}}%
      \put(682,1383){\makebox(0,0)[r]{\strut{}$5$}}%
      \put(682,2061){\makebox(0,0)[r]{\strut{}$10$}}%
      \put(682,2740){\makebox(0,0)[r]{\strut{}$15$}}%
      \put(682,3418){\makebox(0,0)[r]{\strut{}$20$}}%
      \put(682,4097){\makebox(0,0)[r]{\strut{}$25$}}%
      \put(682,4775){\makebox(0,0)[r]{\strut{}$30$}}%
      \put(814,484){\makebox(0,0){\strut{}$0.01$}}%
      \put(1812,484){\makebox(0,0){\strut{}$0.1$}}%
      \put(2810,484){\makebox(0,0){\strut{}$1$}}%
      \put(3809,484){\makebox(0,0){\strut{}$10$}}%
      \put(4807,484){\makebox(0,0){\strut{}$100$}}%
      \put(5805,484){\makebox(0,0){\strut{}$1000$}}%
      \put(6803,484){\makebox(0,0){\strut{}$10000$}}%
    }%
    \gplgaddtomacro\gplfronttext{%
      \csname LTb\endcsname%
      \put(176,2739){\rotatebox{-270}{\makebox(0,0){\large $\widetilde{u}^+$}}}%
      \put(3808,154){\makebox(0,0){\large $z^+$}}%
      \csname LTb\endcsname%
      \put(5816,1317){\makebox(0,0)[r]{\strut{}LU model}}%
      \csname LTb\endcsname%
      \put(5816,1097){\makebox(0,0)[r]{\strut{}DNS}}%
      \csname LTb\endcsname%
      \put(5816,877){\makebox(0,0)[r]{\strut{}Classical Laws}}%
    }%
    \gplbacktext
    \put(0,0){\includegraphics{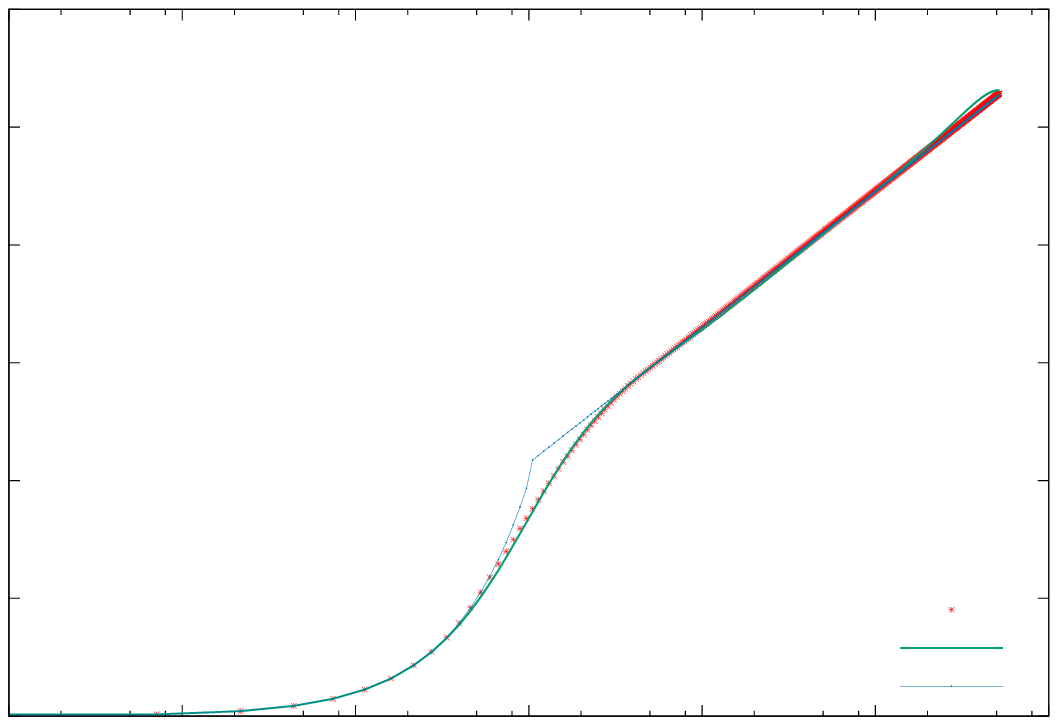}}%
    \gplfronttext
  \end{picture}%
\endgroup